\documentclass[12pt]{article}
\usepackage{amsfonts,latexsym}
\usepackage[dvips]{graphicx}
\usepackage{epsfig}
\usepackage{html}
\usepackage{lscape}
\usepackage{html}
\usepackage{colordvi}
\font\mybb=msbm10 at 12pt
\def\bb#1{\hbox{\mybb#1}}

\def\R {\bb{R}}
\def\unit{\hbox to 3.3pt{\hskip1.3pt \vrule height 7pt width .4pt \hskip.7pt
\vrule height 7.85pt width .4pt \kern-2.4pt
\hrulefill \kern-3pt
\raise 4pt\hbox{\char'40}}}

\def\half{{\textstyle {1 \over 2}}}

\newcommand{\be}[3]{\begin{equation}  \label{#1#2#3}}

\newcommand{\ee}{ \end{equation}}
\newcommand{\ba}{\begin{array}}
\newcommand{\ea}{\end{array}}

\renewcommand{\arraystretch}{1.7}
\font\mybb=msbm10 at 12pt
\def\bb#1{\hbox{\mybb#1}}
 
 \def\R {R}
 \def\E {\bb{E}}

\def\half{{\textstyle {1 \over 2}}}

\begin{document}

\begin{flushright}
\small
UG/2--99\\
HUB-EP-99/12\\
{\bf hep-th/9907006}\\
July $1$st, $1999$
\normalsize
\end{flushright}

\begin{center}

%title

\vspace{.7cm}

{\huge {\bf On Domain--wall/QFT Dualities

in Various Dimensions}}

\vspace{1.7cm}

%authors

{\bf\large Klaus Behrndt}${}^{\spadesuit}$
\footnote{E-mail: {\tt behrndt@qft2.physik.hu-berlin.de},
{\tt behrndt@theory.caltech.edu}\\ 
\hspace*{0.5cm} Present address: {\it California Institute of Technology, Pasadena, CA 91125, USA}},
{\bf\large Eric Bergshoeff}${}^{\diamondsuit}$
\footnote{E-mail: {\tt E.Bergshoeff@phys.rug.nl}},
{\bf\large Rein Halbersma}${}^{\diamondsuit}$
\footnote{E-mail: {\tt R.Halbersma@phys.rug.nl}}\\
{\bf\large and Jan Pieter van der Schaar}${}^{\diamondsuit}$
\footnote{E-mail: {\tt schaar@phys.rug.nl}}\vskip 0.4cm

${}^{\spadesuit}$\ {\it Institut f\"ur Physik, Invalidenstra\ss{}e 110, \\
10115 Berlin, Germany}
\vskip 0.2cm

${}^{\diamondsuit}$\ {\it Institute for Theoretical Physics, Nijenborgh 4,
9747 AG Groningen\\ The Netherlands}

\vspace{1.7cm}

%%%%%%%%%%%%%%%%%%%%%%%%%%%%%%%%%%%%%%%%%%%%%%%%%%%%%%%%%%%%%%%%%%%%%%

{\bf Abstract}

\end{center}

\begin{quotation}

\small

We investigate domain-wall/quantum field theory
correspondences in various dimensions.
Our general analysis does not only cover the well--studied cases 
in ten and eleven
dimensions but also enables us to discuss new cases like 
a Type I/Heterotic 6--brane in ten dimensions
and domain-wall dualities
in lower than ten dimensions. The examples we discuss
include `d--branes' in six dimensions
preserving 8 supersymmetries and extreme black holes
in various dimensions. In the latter case we construct the 
quantum mechanics 
Hamiltonian and discuss several limits. 
\end{quotation}

\newpage

\pagestyle{plain}

%%%%%%%%%%%%%%%%%%%%%%%%%%%%%%%%%%%%%%%%%%%%%%%%%%%%%%%%%%%%%%

\section*{Introduction}

%%%%%%%%%%%%%%%%%%%%%%%%%%%%%%%%%%%%%%%%%%%%%%%%%%%%%%%%%

Anti-de Sitter (AdS) gravity has attracted much attention due to the
conjectured correspondence to a conformal field theory
(CFT) on the boundary of the AdS spacetime  \cite{malda} (for a review, see 
\cite{review}). This CFT is given by the worldvolume
field theory, in the limit that gravity is decoupled, of a brane moving in the
AdS background. In order for this
correspondence to hold, it is essential to have an AdS space, whose isometry
group acts as the conformal group on the boundary \cite{CKP,CKP2}. The regular
$p$--branes (i.e. the D3--brane in $D=10$ dimensions and the M2/M5--branes
in $D=11$ dimensions)
exhibit indeed an AdS spacetime in the so-called near-horizon limit.
However,
for the other branes both the string as well as the Einstein metric do
not lead to a similar anti-de Sitter geometry. 

On the other hand, for all supersymmetric $p$--branes there does 
exist a special frame in
which the $p$--brane geometry, in the near-horizon limit, 
factorizes into a sphere
times an AdS or flat spacetime \cite{9405124,BPS}. Moreover, the near-horizon
geometry of these branes contains a non-trivial dilaton. Choosing a 
particular radial coordinate, the dilaton field is in fact linear in 
that radial coordinate of AdS spacetime.  
The metric of
this special frame is conformally equivalent to the string frame metric and
coincides with the so-called sigma-model metric coupling to a {\em
dual} brane probe. It is therefore called the `{\em dual}
frame' metric \cite{9405124}.  
AdS spaces
can be viewed as special cases (in the sense of having a zero dilaton) 
of AdS spaces with a linear dilaton.
AdS spaces with a linear dilaton are in fact conformally equivalent to 
domain-wall (DW) spacetimes
\cite{9607164}. The presence of the dilaton breaks the scale invariance
and the AdS isometry group
gets reduced to the Poincare isometry group of the DW.

Since all branes are related by duality, it is natural to conjecture
that the AdS/CFT dualities based on the regular $p$--branes can be
generalized to so-called domain-wall dualities, corresponding
to all the other $p$--branes,  where the CFT is
replaced by an ordinary QFT. 
Indeed, for the D$p$--branes ($p= 0, \cdots ,6)$ in $D=10$ dimensions such 
DW/QFT dualities have been discussed in \cite{IMSY, BST}.
Due to the holographic property of an AdS space or
of a flat spacetime with linear dilaton \cite{KS},
the dual frame is the natural basis for a discussion of this
DW/QFT correspondence \cite{BST}.
We note that the QFT is not conformally or scale invariant. There will be a
non-trivial renormalization group flow and only at fixed points (in the
UV and/or IR) can we expect a restoration of the conformal symmetry.
Examples of such fixed points are the limits where
D$p$--branes decompactify to M--branes \cite{IMSY, pepo}.

The purpose of this paper is to extend the discussion of the DW/QFT
correspondence, given in \cite{IMSY, BST} for D$p$--branes in ten dimensions,
to general two-block $p$--branes in various dimensions. All the 
$p$--branes we consider  
have a near-horizon geometry of the form DW $\times$ S
(domain-wall times sphere). The $p$--branes in $D<10$ dimensions
can be viewed as $p$--branes in $D=10$ (or intersections of such
branes with the harmonics identified) reduced over some compact 
manifold K. From the $D=10$ point of view these $p$--branes have
a near-horizon geometry of the form DW $\times$ S $\times$ K.
More general near-horizon geometries of the form
DW $\times$ S $\times$ S $\times$ {\E} (domain-wall times sphere times
sphere times euclidean space) can be obtained by
considering intersections where the harmonics of the participating branes 
have not been identified. Such intersections
and their near-horizon geometries have been studied in the second reference of
\cite{BPS} and in \cite{TC}.

We will start our analysis with a general discussion applicable to any
$p$--brane with a non-zero dilaton. We discuss the
so-called field theory limit, i.e. the low-energy limit in which
gravity decouples from the worldvolume QFT. This limit is defined such that
the  QFT contains at least one fixed coupling constant.
Our analysis shows that, in order to obtain a well-behaved DW/QFT 
correspondence, we need to restrict ourselves to a subclass of $p$--branes.
Our choice of restriction is motivated by the condition that
the so-called
holographic energy (or supergravity probe) scale $u$ is related to the
string energy (or D--brane probe) scale $U$ via fixed quantities only.
The exact relation can be found in equation (\ref{uU}).
This condition leads to two constraints, whose precise form can be found
in (\ref{sufcon}). These constraints tell us to which $p$--branes 
we should restrict ourselves and which coupling constant 
in the worldvolume field theory, i.e. corresponding
to scalars or vectors, etc., should be kept fixed.
We find that these constraints are sufficient to obtain well-behaved
DW/QFT dualities.
The restricted class of $p$--branes is distinguished by the fact that their
effective tension is proportional to the inverse 
string coupling constant. In fact, the ones in $D<10$ can be interpreted
as intersections of D--branes in $D=10$ which are reduced over  
relative transverse directions only. The branes participating
in these intersections are equal in number and
delocalized in the compact relative transverse directions.

We will discuss in more detail various special
cases in ten as well as in lower dimensions. These special cases
include a Type I/Heterotic 6--brane in ten dimensions,
the D8--brane, the so-called `d--branes'
in six dimensions and the extreme black holes in various
dimensions. In this paper we will not discuss the
details of the worldvolume theory, all that we need is the worldvolume
field content. An exception is made for the 
extreme black holes where we will present
the (generalized conformal) quantum mechanics Hamiltonian and 
discuss several limits.

The paper
is organized as follows. In Section 1 we summarize the relationship
between DW and AdS spacetimes.  In Section 2 we calculate
the near-horizon geometries of a generic $p$--brane. In Subsection 2.1
we show that for all $p$
the near-horizon geometries factorize into a DW times sphere geometry
using the dual frame metric.  In the
next Subsection we consider the reduction over the spherical part of the
near-horizon geometry and relate the DW times sphere near-horizon
geometry to a DW spacetime after reduction. 
In a third Subsection we discuss the special
cases in which the DW part of the near-horizon geometry 
becomes conformally flat or pure AdS (with zero dilaton).
In Section 3 we discuss the field theory limit for the general case.
The formulae we give in this Section are applied
to the special cases we discuss in the remaining Sections.
First, in Section 4 we  discuss the 
ten-dimensional D$p$--branes, including the D8--brane,
and the so-called six-dimensional `d$p$--branes' preserving 8 supersymmetries. 
Next, in Section 5 we discuss the quantum mechanics of 0--branes, or
extreme black holes, in various dimensions and in
Section 6 we investigate the special case of a Type I/Heterotic
6--brane in ten dimensions.  
Section 6 is a bit special in the sense that it can be read
independently of the rest of the
paper.  In Section 7 we give our conclusions.  We have included
3 Appendices. In Appendix A we give our $p$--brane charge conventions;
in Appendix B we
elaborate on the dual frame metric and show that it is identical to
the sigma-model metric provided that one uses the sigma model of the
dual brane.  Finally, Appendix C contains a a glossary of
most of the symbols used
in this paper together with a short description and the equation where
the symbol is first used.

%%%%%%%%%%%%%%%%%%%%%%%%%%%%%%%%%%%%%%%%%%%%%%%%%%%%%

\section{Domain-Walls and anti-de Sitter spacetimes}

%%%%%%%%%%%%%%%%%%%%%%%%%%%%%%%%%%%%%%%%%%%%%%%%%%%%%%%%%%%
Before we discuss in the next Section the near-horizon geometries of a
general $p$--brane,
we will first discuss in this Section 
the geometries of domain-wall 
(DW)\footnote{For earlier work on domain wall solutions in d=4
supergravity theories, see \cite{cvetic}.} and anti-de Sitter 
(AdS) spacetimes. It is known that
AdS spaces are special cases
of domain-wall spaces characterised by the absence of a
nontrivial dilaton background \cite{9607164}. We will show 
this explicitly in this section.

Domain-wall spaces solve the equations of motion obtained by varying a
(super)gravity action with a cosmological constant $\Lambda$ and a
dilaton\footnote{We use conventions in which our metric is mostly plus
and the string coupling constant $g_s$ is defined 
as $g_s = e^{\phi(\infty)}$.  The used conventions imply that when
$\Lambda>0$ we have an AdS vacuum and when $\Lambda<0$ we have a de
Sitter vacuum (neglecting the dilaton).}. They correspond to
$p$--branes with worldvolume dimension $d=p+1$ which is one less than
the dimension $D$ of the target spacetime they live in. 
The part of the supergravity action needed to describe the DW solution 
is given by (we use the Einstein frame):

\be100
S^E_{d+1} =  \int {\rm d}^{d+1} x \; \sqrt{g} \Big[ R - {4 \over d-1} 
(\partial \phi)^2 + e^{-b \phi}  \Lambda \Big]\, ,
\ee
where $b$ is an arbitrary dilaton coupling parameter. 

Performing a Poincar\'e dualization, which replaces the cosmological
constant $\Lambda$ by a ($d+1$)-form field strength $F_{d+1}$, 
allows us to discuss
naturally objects of codimension one coupling to a $d$-form potential,
defining a domain-wall.  In terms of the field strength $F_{d+1}$
the action is given by:
\be101
S^E_{d+1}=  \int {\rm d}^{d+1} x \; \sqrt{g} \Big[ R - {4 \over d-1} 
(\partial \phi)^2 -{1 \over 2 (d+1)!} e^{b \phi} F_{d+1}^2 \Big]\, .
\ee

The equations of motion following from the action (\ref{101})
can be solved
using the general $p$--brane Ansatz involving harmonic functions. 
The solutions are (in the Einstein frame)\footnote{In this Section we take
$g_s= 1$. Later, when we discuss the field 
theory limit in Section 3, we will put $g_s$ back into 
the dilaton background expression.}

\begin{eqnarray}
\label{domainwall}
{\rm d}s^2_E &=& H^{- {4 \varepsilon \over (d-1) \Delta_{DW}}} {\rm d}x_{d}^2 +
H^{{-4 \varepsilon d \over (d-1) \Delta_{DW}}  
-2 (\varepsilon +1)} {\rm d}y^2\, , \nonumber \\ 
e^{\phi} &=& H^{-(d-1) b \varepsilon \over 4 \Delta_{DW}}\, ,\\
F_{01 \ldots d-1\,y}&=& \epsilon_{01 \ldots d-1}\, 
\partial_y H^\varepsilon\, , \nonumber 
\end{eqnarray}
with $\varepsilon$ an arbitrary parameter and the parameter
$\Delta_{DW}$  defined by

\begin{equation}
\label{deldw}
\Delta_{DW} = {(d-1) b^2 \over 8} - { 2 d \over d-1}\, .
\end{equation}
This $\Delta_{DW}$ is a useful parameter since it is invariant under reductions
or oxidations (in the Einstein frame) \cite{Pope}\footnote{
Our definition of $\Delta_{DW}$
differs from the definition in e.g. \cite{Pope} (except when $D=10$). The
reason for this is the fact we use different normalizations 
(except when $D=10$) for the
dilaton. For instance,  in \cite{Pope} there is in any dimension
a $\half$ in front of the dilaton kinetic term (using the Einstein frame).
This differs
from our normalization of the dilaton kinetic term, as can be seen from
eq. (\ref{100}).}.

The function $H$ is harmonic on the 1-dimensional transverse space
with coordinate $y$:

\begin{equation}
H=c+Q|y|\, ,
\end{equation}
with $c, Q$ constant. Here it is understood that the domain-wall is positioned
at the discontinuity $y=0$, and for $y<0$ we are allowed to use a 
different value of $Q$. The value of $Q$ can be expressed in terms of a 
mass parameter $m$ in the following way, 
\begin{equation}
Q \varepsilon=m \, ,
\end{equation} 
where $m$ is related to the cosmological
constant through the equation
\begin{equation}
\label{cosmo}
\Lambda={-2 m^2 \over \Delta_{DW}}\, . 
\end{equation}

We see that using the Ansatz (\ref{domainwall}) allows
for an undetermined parameter $\varepsilon$ in the domain-wall 
solution. Note that the charge $Q$ can not be considered a physical
parameter because of its dependence on $\varepsilon$.
 The origin of this parameter is the
fact that there are coordinate transformations, labeled by 
$\varepsilon$, that keep the solution within the same Ansatz. The explicit
form of these coordinate transformations is given in \cite{on9branes}.

Different choices of $\varepsilon$ lead to different expressions for the
metric. For instance, one can always choose $\varepsilon$ such that the metric
is conformally flat:
\begin{equation}
\varepsilon={-\Delta_{DW} \over \Delta_{DW}+2}\, .
\end{equation}
This choice was made in \cite{PW} for the domain-wall in $d+1 = 10$ dimensions.
Another possibility is to choose a `D--brane basis' in which the two powers
of the harmonic functions occurring in the string frame metric are opposite:
\begin{equation}
\varepsilon={-\Delta_{DW}(d-1)  \over b (d-1) + \Delta_{DW} (d-1) + 
2 (d+1)}\, .
\end{equation}
This choice was made in \cite{BGPT} for $d+1 = 10$ dimensions.

To adapt the discussion to what we will encounter in the next Section
when we discuss the near-horizon 
geometries, we shift the position of the domain 
wall to infinity, allowing us to discard the constant $c$ in the 
harmonic function.
Furthermore, we 
get rid of the free parameter $\varepsilon$ by making the following
$y \rightarrow \lambda$ coordinate transformation:
\begin{equation}
Qy =e^{-Q\lambda}\, .
\end{equation}

The domain-wall solution in the new $\lambda$ coordinate reads
\be104
\label{dwgen}
{\rm d}s^2_E = e^{m \lambda \left( (d-1)b^2 \over 4\Delta_{DW} \right)}  
\left( e^{-2m \lambda 
({2+\Delta_{DW} \over \Delta_{DW}})} {\rm d}x_{d}^2 + {\rm d}\lambda^2 \right) 
\quad , \quad e^{\phi} = e^{(d-1) b m \lambda \over 4 \Delta_{DW}}\, .
\ee 
This solution corresponds to the action (\ref{100}) with $\Lambda$ given by
(\ref{cosmo}).

Written like this, it is clear that we can 
perform a conformal transformation to get rid of the
conformal factor in front of the metric. We thus obtain the
regular `dual frame' metric $g_D$ 

\begin{equation}
\label{reg}
g_{D}=e^{-b \phi} g_{E} \, ,
\end{equation}
with
\begin{equation}
\label{dwr}
{\rm d}s^2_D = 
 e^{-2m \lambda 
({2+\Delta_{DW} \over \Delta_{DW}})} {\rm d}x_{d}^2 + {\rm d}\lambda^2 
\quad , \quad e^{\phi} = e^{(d-1) b m \lambda \over 4 \Delta_{DW}}\, .
\end{equation}

Note that when $\Delta_{DW}=-2$ the metric becomes conformally flat.

In the dual frame (\ref{reg}) the domain-wall metric (\ref{dwr})
describes an ${\rm AdS}_{d+1}$ space. In fact this metric has constant
negative curvature, which however is not fixed by the cosmological
constant. The deviation is parameterized by the free dilaton coupling
parameter  $b$
corresponding to a non-vanishing dilaton charge. Furthermore, as discussed
in \cite{9405124}, the linear dilaton corresponds to a non-vanishing
conformal Killing vector, which means that conformal transformations
have to be accompanied by a shift in the dilaton, i.e. yield a running
coupling constant.

%%%%%%%%%%%%%%%%%%%%%%%%%%%%%%%%%%%%%%%%%%%%%%%%%%%%%

\section{Near-horizon Geometries of $p$--branes}

In this Section we discuss the near-horizon geometry of a generic $p$--brane.
In Subsection 2.1 we calculate the near-horizon geometry in terms of the
dual frame metric. In Subsection 2.2 we discuss the reduction over the
spherical part of the near-horizon metric. Finally, in Subsection 2.3
we discuss as special cases the conformally flat and AdS near-horizon
geometries.

\subsection{$p$--branes in the dual frame}

%%%%%%%%%%%%%%%%%%%%%%%%%%%%%%%%%%%%%%%%%%%%%%%%%%%%%%%%%%%
% In this Section we determine the near-horizon geometries of the
% standard branes i.e. the branes that are described by 
% `two-block' supergravity solutions
% (see eq. (\ref{2block}) below) in $D$ dimensions.
Our starting point is the $D$-dimensional action
\be105
S_D = \int {\rm d}^D x \; \sqrt{g} \Big[ R - {4 \over D-2} (\partial \phi)^2 -
{1 \over 2 (\tilde d +1 )!} e^{-a \phi} F_{\tilde d +1}^2 \Big]\, ,
\ee
which contains three independent parameters: the target spacetime 
dimension $D$, the dilaton coupling parameter $a$ and a parameter $p$
specifying the rank $D-p-2$ of the  field strength $F$. 
We have furthermore
introduced two useful dependent parameters $d$ and $\tilde d$
which are defined by 
\begin{equation}
\left\{
\begin{array}{rcl}
d &=& p+1 \ \ \hskip .9truecm  {\rm dimension\ of\ the\ worldvolume}\, , \\
\tilde d &=& D- d -2  \ \  {\rm dimension\ of\ the\ dual\ brane\
worldvolume}\, .
\end{array}
\right. \\
\end{equation}
We next consider the following class of diagonal ``two-block''
$p$--brane solutions (using the Einstein frame)\footnote{We work with 
{\it magnetic} potentials.}:
\begin{eqnarray}
\label{2block}
{\rm d}s^2_E &=& H^{- {4\tilde d \over (D-2) \Delta}} {\rm d}x_{d}^2 +
H^{4d \over (D-2) \Delta} {\rm d}x_{\tilde d + 2}^2\, ,\nonumber\\
e^{\phi} &=& H^{(D-2) a \over 4 \Delta}\, , \\
F &=& ^*( {\rm d}H\wedge {\rm d}x_{1} \wedge \cdots \wedge 
{\rm d}x_{d})\nonumber\, ,
\end{eqnarray}
where $^*$ is the Hodge operator on $D$-dimensional spacetime and
$\Delta$ a generalization of the $\Delta_{DW}$ in the
previous section defined by\footnote{We always have $\Delta >0$ except 
for domain-walls $(\tilde d = -1)$ when $\Delta$ can be negative. 
Note that $\Delta$ is invariant
under toroidal reduction but not under sphere reduction. In Subsection 2.2
we will perform such sphere reductions and obtain domain-walls with
$\Delta_{\rm DW} <0$ out of branes with $\Delta >0$.}
\begin{equation}
\label{delta}
\Delta = {(D-2) a^2 \over 8} + { 2d \tilde d \over D-2}\, ,
\end{equation}
which is, in contrast to the dilaton coupling parameter $a$,
invariant under reductions
and oxidations (in the Einstein frame).
The function $H$ is harmonic over the 
$\tilde d + 2$ transverse coordinates and,
assuming that 
\be115
{\tilde d} \ne -2, 0\, ,
\ee
(i.e. no constant or logarithmic harmonic) 
this harmonic function is given by
\be120
H = 1 + \Big({r_0 \over r}\Big)^{\tilde d}\, ,
\ee
where $r_0$ is an arbitrary integration constant with
the dimensions of length, which is of course related to the
the charge (and mass) of the $p$--brane (see appendix A).

The two-block solutions (\ref{2block}) include 
the (supersymmetric) domain-wall spaces
of the previous section.
They correspond to the case $\tilde d=-1$, $\epsilon=-1$ and $r_0 = 1/m$.
The solutions also include the known branes in ten and eleven dimensions
as well as branes in lower dimensions. In Table 1 we give the values
of $a$ and $\Delta$ of a few cases, including
the well-known cases of the M2--brane, the M5--brane, the 
D$p$--branes 
$(p=0,1,\ldots 6,8,9)$, the fundamental string, or NS1--brane, the
solitonic 5--brane, or NS5--brane, as well as
the `d--branes' 
in six dimensions and the (supersymmetric) dilatonic black holes in 
four dimensions.
For these cases the parameters $a$ 
and $\Delta$ are given by:

{\small
\begin{center}
\begin{tabular}{|c|c|c|c|}
\hline
Dimension&a&$\Delta$&Name\\
\hline
11&0&4&M--branes\cr
10&${3-p\over 2}$&4&D$p$--branes\cr
&--1&4&NS1--brane\cr
&1&4&NS5--brane\cr
6&${1-p}$&2&d$p$--branes\cr
5&0&4/3&m--branes\cr
4&$2 \sqrt 3$&4& black hole\cr
&2&2& ,,\cr
&$2/\sqrt 3$&4/3& ,,\cr
&0&1&RN black hole\cr
\hline
\end{tabular}
\end{center}
\noindent {\bf Table 1}: The Table indicates the values of $a$ and $\Delta$
(in the Einstein frame) for a variety of branes in diverse dimensions.
\bigskip
}

If the branes under consideration preserve any supersymmetries
we can set \cite{Pope}
\be125
\Delta = {4 \over n}\, ,
\ee
where generically $32/{2^n}$ is the number of unbroken supersymmetries.
In case of intersecting branes embedded in a
theory with 32 supersymmetries $n$ is the number
of branes that participate in the intersection\footnote{Sometimes 
branes can be added without breaking more supersymmetries, in which 
case the above rule about the number of unbroken supersymmetries
and the number of intersecting branes involved does not hold.} 

The two-block solutions (\ref{2block}) do {\it not} involve waves and
Kaluza--Klein monopoles. We also excluded D--instantons, whose
near-horizon geometry has been discussed in \cite{Be1, Sk1}. 
The case ${\tilde d} = 0$ that is excluded by (\ref{115}) corresponds to 
branes with 2 transverse directions, e.g. the D7--brane
in $D=10$. 
The other case ${\tilde d}
= -2 $, excluded by (\ref{115}), corresponds 
to spacetime--filling branes. Such branes are given by a flat Minkowski
spacetime solution and have no near-horizon geometries. 

We next consider the limit for which the constant part in the harmonic 
function is negligible, i.e. for the limits
\begin{eqnarray}
\label{nhl}
{r \over r_0 } &\rightarrow& \infty\hskip .6truecm {\rm for}\ \ 
{\tilde d} = -1\, ,
\nonumber\\ 
{r \over r_0}  &\rightarrow& 0 \hskip .8truecm {\rm all\ other\ cases}\, .
\end{eqnarray}
Assuming that the branes are positioned at $r=0$ this limit brings
us close to the brane when $\tilde d>0$. When $\tilde d=-1$ however
this limit takes us far away from the brane. We refer
to this limit as the near-horizon limit.

In this limit the metric and dilaton becomes
\be130
{\rm d}s^2_E = 
\Big({r_0 \over r}\Big)^{- {4 \tilde d^2 \over \Delta(D-2)}} {\rm d}x_{d}^2 
+ \Big({r_0 \over r}\Big)^{4 d  \tilde d \over \Delta(D-2)} 
{\rm d}x_{\tilde d + 2}^2 
\quad , \quad  e^{\phi} = \Big({r_0 \over r}\Big)^
{(D-2) a  \tilde d \over 4 \Delta }\, . 
\ee
As in the previous section we go to the dual frame, 
which is defined by the conformal
rescaling 
\be160
g_D = e^{({a \over \tilde d})\phi}  g_E\, .
\ee
In this dual frame the spherical part factors off in the near-horizon limit
like in \cite{BST}. 

After the transition of the
Einstein metric to the regular `dual frame' metric in the action (\ref{105})
all terms in the action are multiplied by the {\it same}
dilaton factor:
\be170
\label{rcl}
S_D =  \int {\rm d}^D x \; \sqrt{g} e^{\delta \phi} \Big[ R + \gamma
(\partial \phi)^2 - {1 \over 2(\tilde d +1)!} F^2_{\tilde d +1} \Big]
\ee
with 
\be175
 \delta = - {(D-2) a \over 2 \tilde d}\, , \hskip .8truecm
\gamma = {D-1 \over D-2} \delta^2 - {4 \over D-2} \, .
\ee

The metric in the dual frame is given by

\be180
\label{start}
{\rm d}s^2_D = \Big({r_0 \over r}\Big)^{2(1 - {2 \tilde d \over \Delta})}
{\rm d}x^2_{d} + \Big({r_0 \over r}\Big)^2 {\rm d}r^2 + r_0^2 {\rm d}
\Omega_{\tilde d + 1}^2 \ .
\ee

Redefining the radius by 

\begin{equation}
\Big({r_0 \over r}\Big) = e^{-\lambda/r_0}\, ,
\end{equation}
we can write the near-horizon metric ($\lambda \rightarrow +\infty$
for domain-walls and $\lambda \rightarrow -\infty$ for all other branes)
as

\be190
{\rm d}s^2_D = e^{-2(1 - {2 \tilde d \over \Delta }) \lambda/r_0 }
{\rm d}x_{d}^2 + {\rm d} \lambda^2 + r_0^2 {\rm d} 
\Omega_{\tilde d + 1}^2\, , \qquad 
\phi = - {(D-2) a \tilde d \over 4 \Delta} \,  {\lambda \over r_0}.
\ee

We see that the spacetime factorizes into a spherical part and a
linear dilaton
part. 

To make contact with other results (e.g. \cite{malda,BST}) 
it is useful to write
the dual frame solution (\ref{start}) as
\begin{equation}
\label{standard}
{\rm d}s^2_D = r_0^2 \left[\left({u \over {\cal R}}\right)^2 {\rm d}x^2_{d} + 
\left({{\cal R} \over u}\right)^2 {\rm d}u^2
+ {\rm d}\Omega_{\tilde d + 1}^2\right] \quad , \quad 
e^{ \phi}=r_0^{-{(D-2)a \over 8} \left ( {\beta +1\over \beta}
\right )} \left({u \over {\cal R}}\right)^{-{(D-2) a \over 8} \left ( 
{\beta +1\over \beta}\right )}\ ,
\end{equation}
where we made the $r \rightarrow u$ coordinate transformation:
\begin{equation}
\label{radii}
{u \over {\cal R}}= {r^{\beta} \over r_0^{\beta + 1}}\, , \quad 
{\cal R} = {1 \over \beta}
\end{equation}
with $\beta$ given by
\begin{equation}
\label{trans}
\beta={2 \tilde d \over \Delta} -1 \, .
\end{equation}
The horospherical coordinate $u$ can be interpreted as 
the energy scale 
describing supergravity probes \cite{pepo}. In the metric (\ref{standard})
the point $u=0$ is a non-singular Killing horizon\footnote{
Strictly speaking, the near-horizon  (\ref{nhl}) does not always 
correspond to the limit $u \rightarrow 0$.
An exception is the 
$D=10$ D6--brane in which case we have $\beta <0$ and
hence the limit (\ref{nhl}) implies the limit $u \rightarrow \infty$.}
 and $u=\infty$ corresponds
to the boundary of $AdS_{d+1}$.

Using the horospherical coordinate $u$ we recognize the metric 
as that of an AdS space times a sphere. Through (\ref{radii}) we see 
that ${\cal R}$ is the size of the AdS part relative to the sphere. 
As expected we 
recover in the cases of D3, M2 and M5 the well known relations between
the radius of the sphere and the radius of AdS, being respectively 
${\cal R}=1$, ${\cal R}=\half$ and ${\cal R}=2$.
The dimensionful length scale ${\cal R} r_0$ is the radius
of the hyperboloid surface embedded in a $p+2$ dimensional flat spacetime 
with $SO(2,p)$ isometry, giving the (induced) metric (\ref{standard}).  

Summarizing, in this Subsection we showed that in the dual frame, defined by
(\ref{160}),  all $p$--branes have a near 
horizon geometry that factorizes into a domain-wall spacetime 
${\rm DW}_{d+1}$
times a 
sphere $S^{\tilde d +1}$. Note that the domain-wall part of the metric
has all the isometries of an AdS space. These isometries are 
broken in the complete background because of a 
nontrivial linear dilaton.

\subsection{Reduction over the sphere}

%%%%%%%%%%%%%%%%%%%%%%%%%%%%%%%%%%%%%%%%%%%%%%%%%%%%%%%%%%%

Sphere reductions
of supergravities that admit domain-wall $\times$ sphere vacuum 
solutions (as opposed to AdS $\times$ sphere vacuum solutions)
have only been recently discussed in the literature \cite{BST,CLLP}.
We expect that 
when we reduce over the $\tilde d + 1$ angular variables of the sphere
we will end
up with a gauged  supergravity in $d+1$ dimensions whose action
contains the terms
\be240
S^R_{d+1} =  \int {\rm d}^{d+1} x \; \sqrt{g} e^{\delta \phi} \Big[ R + \gamma 
(\partial \phi)^2
+ \Lambda \Big]\, .
\ee
The number of supersymmetries of this action is determined by the original
$p$--brane. The value of $\Lambda$ can be determined from the 
$\tilde d +1$-form curvature which, in the magnetic picture, 
is just the volume form 
on the sphere and there is a contribution from the Ricci scalar.
We will give the value of $\Lambda$ after 
we determined the domain-wall solution in the Einstein frame.
Transforming to the Einstein frame\footnote{This transformation 
to the Einstein frame cannot be performed for $p=0$.} 
\be250
{\rm d}s_E^2 = e^{{2\delta\over (d-1)}\phi} {\rm d}s^2_D,
\ee
and rescaling 
the dilaton as $\phi \rightarrow \phi / c$ with 
(the parameter $\beta$ is defined in (\ref{trans}))
\be252
c^2  = {(\beta +1 ) \tilde d \over \tilde d - \beta} \, ,
\ee
we get for the action {\em exactly} the expression given in 
(\ref{100}), with dilaton coupling parameter $b$ given by 
\be270
b = a \, {(D-2) d \over (d-1) \tilde d} \, c \, .
\ee
The worldvolume part of the solution in this Einstein frame is given by
\be285
\label{DWd+1}
{{\rm d}s_E}^2 = e^{ b^2 (d-1)(\tilde d -\beta) \lambda/ 8r_0 } 
\Big[ e^{2\beta \lambda/r_0 } {\rm d}x_{d}^2 + {\rm d}\lambda^2 \Big]
\qquad, \qquad 
  \phi = - {(D-2) a \tilde d \over 4 \Delta}\,  {\lambda \over r_0 c}\, .
\ee

For the generic case this is a domain-wall solution ${\rm DW}_{d+1}$
in $d+1$ spacetime dimensions and in order to relate it to the ones discussed
in Section 1, we must give the parameter $\Delta_{DW}$. We find that 
$\Delta_{DW}$  is given by
\be286
\label{deldwp}
\Delta_{DW} = \frac{-2\tilde d}{(\tilde d - \beta)}\, .
\ee
Using the above expression and identifying 
\begin{equation}
m={-\tilde d \over r_0}\, ,
\end{equation}
we recover the general form for the metric given by (\ref{dwgen}).
We can also use the general analysis in Section 1 to determine
the value of the cosmological constant $\Lambda$ given in (\ref{cosmo}).
We find
\begin{equation}
\Lambda = {\tilde d \over 2r_0^2} \left[ 2(\tilde d+1)-{4\tilde d \over \Delta}
\right] \, .
\end{equation}
The reduction of the Ricci scalar is responsible for the first term 
in this expression and the second term is from the reduction of
the (magnetic) $(\tilde d+1)$-form curvature. Analysing this
expression we find that all $p$--brane near-horizon geometries
give a $\Lambda>0$ (with $1 \leq \tilde d \leq (D-3)$) except
for the domain-walls, which have $\tilde d=-1$ and a sign change
occurs, giving $\Lambda<0$. We note that this is not in contradiction
with the fact that all $p$--branes (including the domain-walls) 
have AdS geometries in the near-horizon limit (as defined in Section 2), 
because the linear dilaton will also contribute to an 
{\it effective} cosmological constant which is always positive.

Notice that the right hand side of (\ref{deldwp}) only depends on 
$\tilde d$. This expresses the
fact that when we perform a double dimensional reduction (under
which $\tilde d$ is invariant), the near-horizon 
geometries of the $p$- and $(p-1)$--brane are also related by a double 
dimensional reduction. In contrast, $\Delta_{DW}$ is not invariant
under a direct dimensional reduction 
($\tilde d \rightarrow \tilde d -1$):
the near-horizon domain-wall solution is 
mapped to another near-horizon domain-wall solution with $\tilde d$ replaced 
by $\tilde d -1$.

For the convenience of the reader we give three Tables in which the 
value of $\Delta_{DW}$
belonging to different $p$--branes are given.
% \bigskip

{\small
\begin{center}
\begin{tabular}{|c||c|c|c|c|c|c|c|c|}
\hline
Brane&D0&{D1/F1}&D2&{D3}&D4&{D5/NS-5}&D6&D8\\
\hline
$\Delta_{\rm DW}$&${\textstyle -{28\over 9}}$&
${\textstyle {-3}}$&
${\textstyle -{20\over 7}}$&
${\textstyle  -{8\over 3}}$&
${\textstyle -{12\over 5}}$&
${\textstyle  {-2}}$&
${\textstyle -{4\over 3}}$&
${\textstyle {4}}$\cr
\hline
\end{tabular}
\end{center}
\noindent {\bf Table 2}: The values of $\Delta_{DW}$
for the different ten dimensional D$p$--branes.
\bigskip
}
\bigskip

{\small
\begin{center}
\begin{tabular}{|c||c|c|c|c|}
\hline
d$p$&d0&{d1}&d2&{d4}\\
\hline
$\Delta_{\rm DW}$&--6&--4&--2&2\cr
\hline
\end{tabular}
\end{center}
\noindent {\bf Table 3}: The values of $\Delta_{DW}$
for the different d$p$--branes in six dimensions with $n=2$.
\bigskip
}
\bigskip

{\small
\begin{center}
\begin{tabular}{|c||c|c|c|c|}
\hline
a&{$2\sqrt 3$}&2&$2/\sqrt 3$& 0\\
\hline
$\Delta_{\rm DW}$&--4/3&--2&-4&$\infty$\cr
\hline
parent&D6 & d2 & m1 & --\cr
\hline
\end{tabular}
\end{center}
\noindent {\bf Table 4}: The values of $\Delta_{DW}$
for the different dilatonic black holes in four dimensions and 
the corresponding `parent' branes in higher dimensions.
\bigskip
}
\bigskip

\subsection{Special Cases}

%%%%%%%%%%%%%%%%%%%%%%%%%%%%%%%%%%%%%%%%%%%%%%%%%%%%%%%%%%%

In this Subsection we discuss two special cases
where the domain-wall ${\rm DW}_{d+1}$ in the near-horizon geometry
becomes a (conformally) flat space
$\mathbb{R}^{d,1}$
or where the dilaton vanishes and the spacetime becomes an
anti-de Sitter spacetime ${\rm AdS}_{d+1}$.
\bigskip

\noindent {\bf Conformally flat spaces}

We require that the near-horizon geometry is a flat space $\mathbb{R}^{d,1}$
in the dual frame.
{From} (\ref{180}) we see that the condition for flat space is given by:
\be290
\label{condition}
2 \tilde d = \Delta\, ,
\ee
which, using (\ref{trans}),  can also be expressed as $\beta=0$.

This implies that
\be291
\Delta_{DW}=-2 \qquad, \qquad b^2=\frac{16}{(d-1)^2}\, ,
\ee
and thus the domain-wall solution ${\rm DW}_{d+1}$ becomes
\be292
{{\rm d}s_E}^2 = e^{\frac{2\tilde d}{d-1}  \lambda/r_0 } 
\left[ {\rm d}x_{d}^2 + {\rm d}\lambda^2 \right]
\qquad, \qquad 
 \phi = {- a\,\lambda \over r_0 c},
\ee
with $a^2=\frac{4 \Delta^2}{(D-2)^2}$.

We now look to solutions of condition (\ref{condition}).
We restrict ourselves to supersymmetric
solutions enabling us to replace $\Delta$ by $4/n$.
For $n=1$ this condition yields
\be300
p = D- 5   \, ,
\ee
which leads to 5--branes in 10 dimensions and all double dimensional
reductions of the 5--brane to lower dimensions. For the next case, $n=2$, 
the condition (\ref{condition}) yields
\be310
p = D - 4  \, ,
\ee
which leads to 6--branes in 10 dimensions and all
double dimensional reductions of this 6--brane in lower dimensions.
 Notice that $32/2^n$ is the number
of unbroken supersymmetries and therefore this must be a Type I or
a Heterotic  6--brane. Such branes are charged 
under Yang-Mills gauge field, which is broken to an Abelian
$U(1)$ gauge field. For the heterotic case the solution in the
string frame metric $g_S$ is given by ($a = 1/2$)

\be320
{\rm d}s^2_H = {\rm d}x_{d}^2 + H^2 {\rm d}x_{\tilde d + 2}^2 \quad , 
\quad e^{2\phi} = H\, ,
\ee

and the type I solution reads ($a = -1/2$)

\be330
{\rm d}s^2_I = H^{-1/2}  {\rm d}x_{d}^2 + H^{3/2} {\rm d}x_{\tilde d + 2}^2 
\quad , \quad e^{-2\phi} = H \ .
\ee

In both cases the solution for the magnetic gauge field coincides 
with the solution for the type II
6--brane gauge field. 
Note, that the non-spherical part in both cases
is not flat, but {\em conformally} flat near the horizon. We will
discuss the Type I/Heterotic 6--brane in more detail in Section 6.
\bigskip

\noindent {\bf Anti-de Sitter spaces}

The anti-de Sitter case ${\rm DW}_{d+1} = {\rm AdS}_{d+1}$ 
corresponds to the special 
case that $a = b = 0$.
This condition yields (see (\ref{delta})) 
\be331
{2 d\tilde d \over \Delta}=  (d+\tilde d)\, .
\ee

It also implies that
\be332
\Delta_{DW}=-2 \frac{d}{d-1} \qquad {\rm and}\qquad
\beta = {\tilde d\over d}\, .
\ee

The solution for these cases becomes 
(the overall conformal factor vanishes)
\be333
{{\rm d}s_E}^2 = e^{2 \beta \lambda/r_0} {\rm d}x_{d}^2 + d \lambda^2 
\qquad, \qquad 
 \phi = 0\, .
\ee

The condition ({\ref{331}}) can be satisfied 
for the cases that we preserve some amount of supersymmetry 
(i.e. using (\ref{125})). We find all the 
well known cases as we have summarized in Table 5 below 
\cite{CKP2,BPS,kaku}.

{\small
\begin{center}
\begin{tabular}{|c|c|c|c|c|}
\hline
$D$ &$\beta$&$\Delta_{\rm DW}$&$\Delta$&Name\\
\hline
11 & $\frac{1}{2}$ & --12/5    & 4   & M5--brane \cr
   &      2        &  --3      & 4   & M2--brane \cr
10 &      1        & --8/3     & 4   & D3--brane \cr
 6 &      1        &  --4      & 2   & d1--brane \cr
 5 & $\frac{1}{2}$ &  --4      & 4/3 & m1--brane \cr
   &      2        & $\infty$ & 4/3 & m0--brane \cr
 4 &      1        & $\infty$ &  1  & RN black hole\cr
\hline
\end{tabular}
\end{center}
\noindent {\bf Table 5}: The Table indicates the values of $\beta$, 
$\Delta_{\rm DW}$ and $\Delta$ (in the Einstein frame) for all 
branes that have an AdS near-horizon geometry.
\bigskip
}

The Table shows that there are branes with an AdS geometry for $n=1, \ldots 4$
and that these branes have only {\em three} different values for 
$\beta$: ${1 \over 2}, 1$ and $2$. Note that the m--branes have the same values
for $\beta$ as the M--branes. In fact, it turns out that the m1--brane is an
intersection of 3 M5--branes and that the m0--brane
an intersection of 3 M2--branes.
A similar thing happens for the d1 and the D3; they both have $\beta=1$
and the d1 is an intersection of 2 D3--branes. 

Note that the case $p=0$ is special:
for $p=0$ the value of $\Delta_{\rm DW}$ blows up and 
using (\ref{cosmo}) we see that the cosmological constant vanishes. 
However, the corresponding ${\rm AdS}_2$ space still satisfies the Einstein
equation in 2 dimensions and has a nonzero curvature.

\section{The field theory limit}

In this Section we first give a general discussion of the field theory
limit of $p$--branes (Subsection 3.1). For the
convenience of the reader we discuss in the following two Subsections, 
as a familiar example, the Pure AdS cases (Subsection 3.2) and the
Conformally flat cases (Subsection 3.3). In Section 4 and 5 we will
discuss examples, some of which are known but others have not been
studied before .

\subsection{The general case}

%%%%%%%%%%%%%%%%%%%%%%%%%%%%%%%%%%%%%%%%%%%%%%%%%%%%%%%%%%%

The fact that every $p$--brane near-horizon metric factorizes 
into a domain-wall times a sphere suggests that we can relate
supergravities on such a background to field theories on the
boundary in the (generalized) way described in \cite{IMSY,BST}.
The conformally flat case requires a special treatment. This case
will be separately discussed in Subsection 3.3.

When discussing the field theory limit, one usually starts by
writing the harmonic function in terms of appropriately defined 
field theory (energy and coupling) parameters, which are kept fixed,
and then shows that taking the limit $\alpha^\prime \rightarrow 0$ 
(low energy limit) the harmonic function tends to infinity, meaning 
we end up in the near-horizon region. This near-horizon metric is then
shown to stay finite in $\alpha^\prime$ units when taking the low 
energy limit.

Because we want to discuss $p$--brane backgrounds in general, we do not 
want to start by defining appropriate field theory variables 
(because in general this will change for every $p$--brane). Instead
we will proceed in the opposite direction. Starting from
the near-horizon geometry, we perform the $r \rightarrow u$ coordinate
transformation given in (\ref{radii}).
In terms of the the new `holographic' coordinate $u$
the near-horizon metric is AdS for all $p$--branes and it is finite 
in $\alpha^\prime$ units. Afterwards we  perform 
a consistency check and show that the low energy limit indeed takes us into 
the near-horizon region.

Until now we assumed $g_s \equiv 1$ for simplicity, but for 
the present discussion
we must re--insert the explicit $g_s$ dependence into the expression
for the dilaton.
In other words we multiply our present expression for the dilaton background
with an extra factor $g_s$ such that $e^{\phi(\infty)} = g_s$. 
Now that we have
an explicit $g_s$ in our dilaton background we have to re--analyse the
conformal transformation that takes us to the dual frame.
We observe that when going to the dual frame via a conformal transformation
we will 
be left with an overall power of $g_s$ in the dual frame metric
(\ref{160}). To avoid this we must redefine the conformal transformation 
with appropriate
powers of $g_s$ (see the equation below).
To obtain a factor of $\alpha^\prime$ in front of the 
metric (\ref{standard}) instead of $r_0^2$ we must make a further
modification of the conformal transformation. Using the fact that 
$\alpha^\prime = \left(r_0 \over L\right)^2$ (with $L$ defined by 
(\ref{numfactor}), see the appendix), we redefine the conformal 
transformation to the dual frame in the following way
\begin{equation}
\label{limdual}
g_D=L^{-2} \left({e^{\phi} \over g_s}\right)^{{a \over \tilde d}} g_E.
\end{equation}
The redefinition (\ref{limdual}) 
of the conformal transformation will introduce 
explicit\footnote{With explicit we mean $g_s$ dependence not hidden 
in $e^{\phi}$.} $g_s$ and $L$ dependence in the dual frame action.
These dependences are determined as follows.
As the low energy limit of string theory,
the string frame solutions have no explicit dependencies on $g_s$. 
In the action the gravitational + dilaton sector scales as the
usual $1/g_s^2$ (hidden in the dilaton). This should not change when we 
go to the Einstein frame. If we choose the metric in the Einstein frame 
to not carry explicit $g_s$ dependence (as we do in this paper),
the gravitational part of the action should scale explicitly with 
$1/g_s^2$ (there is no dilaton to hide it in). The same holds for the 
kinetic antisymmetric tensor part in the action. The (implicit) scaling with
$g_s$ can be read off by looking at the dilaton coupling in the string 
frame. This will then determine the (explicit) scaling with $g_s$ of the
kinetic antisymmetric tensor part in the Einstein frame. The result
is the following action in the Einstein frame
\begin{equation}
\label{einst}
S_E =  \int {\rm d}^D x \; \sqrt{g} \left\{ {1 \over g_s^2} \Big( R -{4 \over
D-2}(\partial \phi)^2 \Big) - {g_s^{{2 (\tilde d- d) \over D-2}}  \over 2 
(\tilde d +1)!} e^{-a \phi} F^2_{\tilde d +1} \right\}.
\end{equation}
Using the conformal transformation (\ref{limdual}), starting
from the action (\ref{einst}), the dual frame action becomes
\begin{equation}
\label{Ngs}
S_D =  \int {\rm d}^D x \; \sqrt{g} \, (d_pN)^{D-2 \over \tilde d}
e^{\delta \phi} \left\{ R + \gamma
(\partial \phi)^2 - {1  \over 2 (d_p N)^2 
(\tilde d +1)!} F^2_{\tilde d +1} \right\} \, .
\end{equation}

The dual frame solution (\ref{standard}) now reads
\begin{equation}
\label{limitsol}
{\rm d}s^2 = \alpha^\prime \left[ \left({u \over {\cal R} }\right)^2 
{\rm d}x^2_{d} + \left({{\cal R} \over u}\right)^2 {\rm d}u^2
+ {\rm d}\Omega_{\tilde d + 1}^2 \right] , \quad 
e^{\phi}= g_s {r_0}^{{-(D-2) a \over 8} \left( {\beta +1 \over \beta} 
\right)}  \left({u \over {\cal R}} \right)^{{-(D-2) a \over 8} \left( 
{\beta +1 \over \beta} 
\right)} \, .  
\end{equation}
This shows that the AdS metric stays constant in $\alpha^\prime$ units
when taking the low energy limit if we keep the parameter $u$ fixed.
It is clear that this analysis does not hold for the conformally
flat cases ($\beta=0$). We will discuss this case separately in
Subsection 3.3.

We want the field theory on the brane to be nontrivial and
decoupled from the bulk supergravity in the limit
$\alpha^\prime \rightarrow 0$ (low energy limit). Nontrivial means that at 
least one coupling constant in the field theory should stay fixed when 
taking this limit. Because generically these coupling constants will be 
dimensionful (depending on $\alpha^\prime$), fixing this coupling
constant will be nontrivial. Based on dimensional analysis and the scaling
of the effective tension with $g_s$ (see Appendix B) 
we can deduce that the $p$--brane worldvolume field theory has
a ('t Hooft) coupling constant 
$g^2_f$ which can be written as follows:
\begin{equation}
\label{gfix}
g_f^2 = c_p N g_s^{k} \Big(\sqrt{\alpha^\prime}\Big)^{x}\, .
\end{equation}
Here $N$ denotes the number of stacked branes and the scaling of the 
coupling constant with $g_s$ is as the inverse tension
${\tau_p}^{-1}$, involving a constant $k$ equal to 
\begin{equation}
\label{gscale}
k = {a \over 2}+{2 d \over D-2}\, .
\end{equation}
The undetermined numerical factor $c_p$ and the parameter $x$ depend
on the specific field theory under consideration. 
Depending on whether $x$ is positive 
or negative, the factor in (\ref{gfix}) 
involving $N$ and $g_s$ should either become large or small in the 
low energy limit in order to 
keep $g_f^2$ fixed.
Decoupling of the bulk supergravity is ensured when the $D$-dimensional
Newton constant vanishes:
\begin{equation}
G_D = {({2\pi l_s})^{(D-2)}\over
32 \pi^2} g_s^2  \rightarrow 0 \, .
\end{equation}
When using the 't Hooft coupling constant $g_f$ as a fixed parameter, 
like we do, we can always
keep $g_s$ small (when $x>0$ we can tune $N \rightarrow \infty$ to fix 
$g_f^2$ instead of taking $g_s \rightarrow \infty$) 
and thus in the low energy limit
$G_D \rightarrow 0$ and (D--dimensional) gravity decouples. 
If we do not use the freedom
to tune $N$, the analysis becomes more subtle because the bulk theory can
become strongly coupled ($g_s \rightarrow \infty$). Sometimes
this implies that the decoupling of gravity must be analyzed
in $D+1$ dimensions and in some cases this decoupling
is impossible \cite{hullim}.
This is for instance the case
 for all domain-wall $p$--branes and the D6--brane in $D=10$.

In general a $p$--brane worldvolume field theory
will involve different multiplets and thus different coupling constants.
These coupling constants only differ in their dependence
on $\alpha^\prime$. Fixing one of these coupling constants will 
ensure that the other coupling constants become either
small or large in the low energy limit. When a coupling constant 
becomes small that sector of the $p$--brane field theory will decouple.
For a large coupling constant it will depend on the $p$--brane 
field theory under consideration what will happen. When considering
Yang-Mills theories with 8 supercharges, the theory will involve 
vector- and hyper-multiplets. The vacuum structure of the theory is
described by two different branches, the Higgs branch where the scalars 
in the hypermultiplet have expectation values and a Coulomb branch where
the scalars in the vector multiplet have expectation values. When we consider
all the branes on top of each other, as we will do, the expectation values 
of the vector multiplet scalars (describing the relative positions
of the stacked branes) are zero. This means the theory can only 
be in the Higgs branch (with non-zero hyper multiplet scalars)
where all the vectors become massive 
(except for the $U(1)$ center of mass 
multiplet describing the motion of all the branes in the transverse space).
The masses of the vectors will be proportional to the vector multiplet
coupling constant and will therefore decouple if we take that coupling 
constant to infinity. We will be left with a theory consisting
of scalars only. Note that we do not specify the vacuum structure of the 
field theory
precisely. This is related to the fact that, when we consider an 
interpretation
in terms of intersecting branes, the participating branes are {\it 
delocalized} in the compact relative transverse directions.
The above sketched scenario applies
when we discuss d$p$--branes in $D=6$ (Subsection 4.2) and has been discussed
extensively in the case of pure $AdS_3$ \cite{malda} and in relation
with matrix models of M-theory on $T^5$ \cite{witten, abkss}.

To summarize, when we speak of the field theory limit we take \footnote{
Instead of $\alpha^\prime \rightarrow 0$ and $u =$ fixed one can also
say that at an energy scale $u$ one takes the limit $\sqrt {\alpha^\prime}
u \rightarrow 0$.}
\begin{equation}
\alpha^\prime \rightarrow 0, \quad u = {\rm fixed}, \quad g^2_f = {\rm fixed}.
\end{equation}
{Furthermore, from} the start 
we assume that $g_s \ll 1$. This is 
necessary in order to make sure that the supergravity approximation is valid.
As a consistency check, we must show now that 
the above limit takes us into the near-horizon
metric. For this purpose we rewrite the harmonic function 
in  terms of the fixed quantities
$u$ and $g_f^2$, leaving a power of $\alpha^\prime$:
\begin{equation}
H=1 + (\sqrt{\alpha^\prime})^{{x - \tilde d \over \beta}} 
(g_s)^{{2(k-1)\over \beta}}\left[g_f^2 \,  
\left({u \over {\cal R}}\right)^{\tilde d} \, 
\left({d_p \over c_p}\right) \right]^{-1/\beta}   \, .
\end{equation}
Since we must have $H\rightarrow \infty$ in the field theory limit, 
it is obvious that the exponents of $\alpha^\prime$ and $g_s$ should not 
{\em both} be positive. We will see in the next sections that in all 
interesting supersymmetric examples ($\beta \neq 0$) we have 
\begin{equation}
\label{consistent}
{x- \tilde d \over \beta} < 0 \, .
\end{equation} 
In some cases the power of $g_s$ is positive, ${2(k-1) \over \beta} > 0$
which could spoil the limiting behaviour of the harmonic function $H$. 
However, we can always tune $g_s$ in such a way that we end up in the near
horizon limit when we take the field theory limit. At the end of this
section we will discuss a restriction on the parameters $x$ and $k$ which
will be sufficient to always satisfy the condition (\ref{consistent}).

Putting all the 
$\alpha^\prime$ dependence of the dilaton background inside the
fixed quantity $g_f^2$ we can write the dilaton background as
\begin{equation}
\label{dilimit}
e^{\phi}= g_s^{1 + {(D-2) a \over 2 \Delta \beta}(k-1)} \left( N g_s^{k}
\right)^{a (D-2) (\tilde d - x) \over 4 \Delta \beta x}
\left[(g_f^{2})^{1/x} \left({u \over {\cal R}} \right)\, 
\left({d_p^{1/ \tilde d} \over c_p^{1/ x}}\right)\right]^{{-(D-2) 
a \over 8} \left({\beta + 1 
\over \beta}\right)}.
\end{equation}
We can trust the background solution when the dilaton is small.
For a positive sign of the overall exponent of $g_s$ in (\ref{dilimit}), 
the dilaton is small everywhere except near the 
point $u=0$ or $u=\infty$, depending on the sign of $-a (\beta+1) / \beta$. 
When the dilaton becomes large we should consider the $S$-dual brane. 
A detailed analysis of where the dilaton is large can not be given in 
this general setup. Note that in general the 
dilaton background explicitly depends on $g_s$.
It turns out that this explicit dependence affects
in a negative way the field theory - 
supergravity dualities we would like to find. 
We will come back to this point at the end of this Subsection
after we have made some
restrictions on the parameters $k$ and $x$.

To trust the supergravity solution requires that the 
curvatures in the string frame remain small. Another way of putting this 
is demanding that the effective string tension (in the dual frame) times 
the characteristic spacetime length is large. {From} the metric
(\ref{limitsol}) we see that the characteristic spacetime length
in the dual frame is of order 1 (the length scales describing the 
radii of the AdS space and the sphere are of order 1 in 
$\alpha^\prime$ units). Calculating the effective string tension in
the dual frame using (\ref{limdual}) we deduce that 
the supergravity approximation will be valid provided
\begin{equation}
\label{sugravalid}
\tau_D = \left( d_p Ne^{(2-k) \phi} \right)
^{2/\tilde d} \gg 1 \, .
\end{equation}

Finally, the perturbative field theory description is valid provided
the effective dimensionless coupling constant $g_{\rm eff}^2$, 
constructed from 
the holographic energy scale $u$ and $g_f^2$, is 
small:
\begin{equation}
\label{perturb}
g^2_{\rm eff} = g^2_f \, u^{x} \ll 1 \, .
\end{equation}
Depending on the sign of $x$ the perturbative field theory description will
either be valid near $u=0$ or $u=\infty$. Note that using the
holographic energy scale $u$ to construct the effective dimensionless
coupling constant means that we use supergravity probes \cite{pepo}.
Also notice that this dimensionless combination of $g_f^2$ and $u$
is the one appearing in the dilaton expression (\ref{dilimit}). 

Summarising, to analyse supergravity - field theory dualities we
should determine the (range of) validity of the supergravity and
field theory descriptions. This leads to the following restrictions.

The supergravity solution can be trusted when
\begin{description}
\item{\bf I.} \ \  The string coupling is small: $g_s=e^\phi \ll 1$ with 
 $e^\phi$ given in eq.~(\ref{dilimit}).

\item{\bf II.} \  The curvatures are small: 
$\tau_D = \left( d_p Ne^{(2-k) \phi} \right)
 ^{2/\tilde d} \gg 1$ with $\tau_D$ given by eq.~(\ref{sugravalid}).
\end{description}
The perturbative field theory is valid when
\begin{description}
\item{\bf III.}
The effective coupling constant $g_{\rm eff}^2$ defined by
(\ref{perturb}) is small, i.e. $g^2_{\rm eff} = g^2_f \, 
u^{x} \ll 1$.
\end{description}

We do not expect that
we can relate supergravity to field theory in a well defined way
for generic values of the parameters $a$, $k$ and $x$. In the
AdS case, corresponding to $a=0$, 
the limit is well defined in the supergravity
background and we can relate supergravity on the AdS background to
a superconformal field theory. This case has been well studied.

Whenever $a\neq 0$, the dilaton background will generically spoil 
the `nice' behaviour in the low energy limit. With `nice' we mean 
that there are finite regions in the supergravity background
where perturbative supergravity is valid, i.e. condition
(\ref{sugravalid}) is satisfied.
For D$p$--branes however we know from the work \cite{malda,BST}
that things work out nicely, so a straightforward proposal would be to 
constrain the analysis to $p$--branes which behave like D$p$--branes in 
$D=10$. For D$p$--branes we know that there exists another energy 
parameter $U=r/\alpha^\prime$, related to the length of stretched
strings, which can be kept fixed in the low energy limit at the same 
time as the holographic energy parameter $u$. This implies that the relation
between $u$ and $U$ only involves fixed quantities. In general
the relation between these two energy parameters is
\begin{equation}
\label{uU}
{u \over {\cal R}} = {\alpha^\prime}^{x+\tilde d -\Delta \over \Delta} \,
g_s^{4 (k-1) \over \Delta} \, \left( \frac{d_p}{c_p} g_f^2 \right)^
{-2 \over \Delta} \,  U^{\beta} \, .
\end{equation} 
We propose to restrict ourselves to those $p$--branes for which
(\ref{uU}) does not involve $\alpha^\prime$ and $g_s$.  
This constrains $k$ and $x$ to be equal to (whenever $a \ne 0$)
\begin{equation}
\label{sufcon}
k=1 \quad , \quad x=\Delta -\tilde d \, .
\end{equation}
Note that the above restrictions are also satisfied for the $a=0$ case
provided that $\beta = 1$. Table 5 shows that these are exactly the cases
where the AdS background is embedded in a string theory. The other
values of $\beta$ ($\beta = {\textstyle {1\over 2}}$ or $\beta = 2$)
correspond to theories without a string coupling constant like M-theory.

The nice thing about the second formula in (\ref{sufcon})
is that it immediately tells
us which coupling constant in the $p$--brane field theory we should
keep fixed. For $D=10$ and $n=1$ we find $x=p-3$, which corresponds to the 
scaling of the Yang-Mills coupling constant. Hence, for D$p$--branes in 
ten dimensions we should keep the Yang-Mills coupling constant fixed.
The constraints (\ref{sufcon}) imply that the 
condition (\ref{consistent}) 
is always satisfied. They also ensure that in the dilaton background
all dependence on $g_s$ disappears:
\begin{equation}
\label{dilback}
e^{\phi}= {1 \over N}
\left[(g_f^{2})^{1/x} \left({u \over {\cal R}} \right)\, 
\left({d_p^{1/ \tilde d} \over c_p^{1/ x}}\right)\right]^{{-(D-2) a \over 8} 
\left({\beta + 1 \over \beta}\right)} \, .
\end{equation}
This expression for the dilaton background will guarantee the existence
of finite regions in the background where perturbative supergravity is
valid, i.e. (\ref{sugravalid}) is satisfied. In fact, in the expression
(\ref{sugravalid}) all $N$-dependence will drop out. 
So we conclude that (\ref{sufcon}) is a {\it sufficient} condition
on $k$ and $x$ for obtaining (well behaved) supergravity - field theory 
dualities whenever $a \neq 0$.
We have not proven that it is also a {\it necessary} condition, but we have
not found any counterexamples either. 
As we should expect, imposing $k=1$ will generically relate 
the $p$--branes 
% we will discuss 
to intersecting D$p$--branes in ten dimensions 
(reduced over the relative transverse directions).

\subsection{Pure AdS cases}

The cases where the near-horizon geometry is a pure anti-de-Sitter spacetime 
($a=0$) are considerably simpler, both from the supergravity point of view as 
well as from the field theory point of view, since the full conformal symmetry 
of the background is left unbroken. The supersymmetric pure AdS cases have
been listed in Table 5, Subsection 2.3.

For these cases 
the solution (\ref{limitsol}) reduces to 
\begin{equation}
\label{pure}
{\rm d}s^2 = \alpha^\prime \left[ \left({u \over {\cal R} }\right)^2 
{\rm d}x^2_{d} + \left({{\cal R} \over u}\right)^2 {\rm d}u^2
+ {\rm d}\Omega_{\tilde d + 1}^2 \right]\, .
\end{equation}
The ('t Hooft) coupling constant $g^2_f$ of the $p$--brane field theory should
be independent of $\alpha^\prime$, this means $x=0$. 

{For} the pure AdS cases the parameters $\beta$ and
$\Delta$ are equal to (see Subsection 2.3)
\begin{equation}
\beta={\tilde d \over d} \quad , \quad \Delta={2 d \tilde d \over d+\tilde d}
\, .
\end{equation}
The harmonic function now becomes
\begin{equation}
H=1 + (l_f)^{-d} \left[g_f^2 \,  
\left({u \over {\cal R}}\right)^{\tilde d} \, 
\left({d_p \over c_p}\right) \right]^{-1/\beta}   \, ,
\end{equation}
where $l_f$ is some fundamental length scale, for backgrounds in a 
string theory this is just $\sqrt{\alpha^\prime}$, but for other
theories, like M-theory, this should be $l_p$ (the Planck length).
Because $d>0$, the low energy limit $l_f \rightarrow 0$ always
takes us into the near-horizon region, defined by $H \rightarrow \infty$.

When embedded in a string theory, we can trust the background solution
when the string coupling and the curvature are small ($g_s \sim 0$ and
$\tau_D \gg 1$). Looking at (\ref{sugravalid}) (taking $k=1$ and
$e^\phi=g_s$) we see that we need
\begin{equation}
N\,g_s \gg 1 \, ,
\end{equation}  
which can only be satisfied for large $N$. When not discussing
embeddings in string theory, we loose the parameter $g_s$ and
we only have to require small curvature, which can again only
be satisfied for large $N$. This then immediately leads us
to the (by now well established) conjecture that large $N$ 
superconformal field theory is dual to supergravity on a (pure) AdS 
background.

%%%%%%%%%%%%%%%%%%%%%%%%%%%%%%%%%%%%%%%%%%%%%%%%%%%%%%%

\subsection{Conformally flat cases}

%%%%%%%%%%%%%%%%%%%%%%%%%%%%%%%%%%%%%%%%%%%%%%%%%%%%%%%%%

In this Subsection we analyse the conformally flat cases.
Conformally flat cases have $\beta=0$, for which the solution
(\ref{limitsol}) no longer is defined. However, the singular $\beta$
dependence can be removed by transforming the holographic energy scale
$u$ to the brane energy scale $U$ using (\ref{uU}). Since this
transformation requires the conditions (\ref{sufcon}) to be satisfied
in order for it to be independent of $\alpha^\prime$ and $g_s$ we will
from now on restrict ourselves to those cases. Combining this with the
fact that $\beta = 0$ yields
\begin{equation}
x = \tilde d = {\Delta \over 2} = {D-2 \over 4} \quad , \quad a=-1.
\end{equation}
In the next Section where we discuss the D$p$--branes ($x=p-3$) and 
`d$p$--branes' ($x=p-1$) we will see that condition (\ref{sufcon})
is indeed satisfied for the conformally flat
cases (corresponding to $p=5$ for D$p$--branes and $p=2$ for `d$p$--branes').
In terms of the energy scale $U$ the solution (\ref{limitsol}) 
with $\beta = 0$ is regular:
\begin{equation}
\label{confflat}
{\rm d}s_D^2 = \alpha^\prime \left[ 
\left(g^2_f {d_p \over c_p} \right)^{-4/\Delta} {\rm d}x_d^2 +
\left( {{\rm d}U \over U} \right)^2 + {\rm d}\Omega_{\tilde d+1}^2
\right]\, .
\end{equation} 
Following \cite{BST}, we make the $U \rightarrow \rho$ coordinate change
\begin{equation}
\rho=\ln \left( g^2_f \, U^{x}
\right)^{1/x}={2\, \ln(g^2_{\rm eff}) \over \Delta},
\end{equation}
where the dimensionless effective coupling constant has now 
been defined using the brane energy scale
\begin{equation}
g^2_{\rm eff} = g^2_f \, U^x.
\end{equation}
Rescaling the worldvolume coordinates by $\left(g^2_f {d_p \over c_p} 
\right)^{2/\Delta}$ and using the dimensionless parameter $\rho$ to 
rewrite the solution (\ref{confflat}) we obtain
\begin{equation}
\label{flat}
{\rm d}s^2_D =\alpha^\prime \left[ {\rm d}x^2_{d} + {\rm d}\rho^2 + 
{\rm d}\Omega_{\tilde d + 1}^2 \right] \quad , \quad \phi= {(D-2) \over 8} \rho
-\, \ln\, (\sqrt{c_p d_p} N)\, ,
\end{equation}
so that the metric has the explicit form of a Minkowski space times a 
sphere: $M_{d+1} \times S_{\tilde d + 1}$.
Expressing the harmonic function in terms of the fixed quantities 
$U$ and $g_f$ we get
\begin{equation}
H=1 + \left(\sqrt{\alpha^\prime}\right)^{-\Delta}  g_f^2 
U^{-\tilde d} \left({d_p \over c_p}\right)  \, .
\end{equation}
and we see that the 
low energy limit takes us into the near-horizon region and our 
analysis is consistent.

We can trust the supergravity background solution when the curvature
and the dilaton are small. Plugging the dilaton into
(\ref{sugravalid}) (recall $k=1$ in this case) we find small curvature
when $\rho \gg 1$.  However in that region, as long as $\tilde d>0$ we
have a large dilaton, suggesting we should turn to the S-dual
description. 
On the other hand the perturbative field theory description is
appropriate when $\rho \ll 1$.  
As long as $\tilde d>0$ we find
the same behaviour for all conformally flat cases.

In the $AdS$ cases the near-horizon `throat', in the low energy limit, 
closes and we end up with a near-horizon
spacetime with a boundary, disconnected from the asymptotic 
region. This does not happen in the conformally flat cases. The 
near-horizon spacetime has no boundary where we can naturally place the 
dual field theory and is not disconnected from the infinite throat region.
Particles in this background can propagate into the throat region,
as opposed to the $AdS$ cases, where the boundary prohibits such effects 
(massive particles can not even reach the boundary). This means the throat
region should be included in the dual description \cite{IMSY}.

\section{Examples}

In this Section we will perform
a more detailed investigation
for some specific, supersymmetric, examples. 
When we have a nontrivial dilaton ($a \neq 0$) our analysis 
will be restricted to $p$--branes with $k=1$ and $x=\Delta-\tilde d$,
for reasons explained in the previous Section.
In Subsection 4.1 we will
first discuss the $D=10$ D$p$--branes, including
the D8--brane. In the next Subsection we will discuss the
d$p$--branes in $D=6$ preserving 
only 8 supersymmetries. In the next Section we will focus our attention on the 
different black holes in $D=10, 6, 5, 4$ and the quantum mechanical
models related to them. 

\subsection{Ten-dimensional D$p$--branes}  

The D$p$--branes, except for the D8--brane,
have already been discussed in \cite{IMSY,BST}. We refer to these papers for
more details. The purpose of this Subsection is to shortly review these
papers using the notation of this paper.

All D$p$--branes, in a low energy limit, are
described by a super Yang-Mills theory. 
The Yang-Mills coupling constant for D$p$--branes in $D=10$ follows from 
expanding the Dirac-Born-Infeld action
\begin{eqnarray}
g^2_{YM} & = & {1 \over (2 \pi \alpha^\prime)^2 \tau_p} \nonumber \\
         & = & 2 \pi g_s (2 \pi l_s)^{p-3} \nonumber \\
         & = & c_p \, g_s l_s^{\,(p-3)},
\end{eqnarray}
where the numerical factor $c_p$ is equal to 
$c_p=(2\pi)^{p-2}$ when considering ten-dimensional D$p$--branes.
The scaling $x$ with $l_s$ is given by $x=p-3=\Delta - \tilde d$, so
the coupling constant we would like to keep fixed is the 't Hooft 
coupling constant $g_f^2 = g^2_{YM} N$: 
\begin{equation}
g_f^2=c_p N g_s \sqrt{\alpha^\prime}^{\,p-3} \, .
\end{equation}
In the low energy limit the scalars in the Yang-Mills theory will
decouple.
Other parameters we need are
\begin{equation}
\Delta=4 \, , \quad \tilde d = 7-p \, ,  \quad \beta=\half (5-p) \, , 
\quad a=\half (3-p) \, .
\end{equation}
As was shown in a more general context ($k=1$ and $x=\Delta -\tilde d$),
the constraint (\ref{consistent}) is satisfied and we are 
able to fix two energy quantities at the same time.
The specific relation between the two energy scales $U$ and $u$
for D$p$--branes is given by
\begin{equation}
U^{5-p} = g_f^2  \left({u \over {\cal R}}\right)^2 \left({d_p \over c_p}\right)
\, .
\end{equation}
We can freely choose between them corresponding to considering 
holographic supergravity probes or D$p$ brane probes. The explicit 
values of the constants $d_p$ and $c_p$ can be found in the appendix.  

To keep $g_f^2$ fixed for $p<3$ we do not need to tune $N$, because 
we always take $g_s$ to be small (for the supergravity solution to 
be valid we need low energy and small coupling). For $p>3$ we have to 
take $N$ to infinity to keep $g_f^2$ fixed. As is well known, in 
the special case that $p=3$ we loose the non-trivial dilaton and obtain
a pure $AdS_5 \times S_5$ background. The case $p=5$ is special because it 
is conformally flat. This case was already discussed in Subsection
3.3. We will not discuss the case $p=7$ corresponding to $\tilde d=0$.
 
Plugging in the different parameters the dilaton background
(\ref{dilimit}) takes the form
\begin{equation}
e^{\phi}= {1\over N}\, \left[(g_f^2)^{1 \over p-3} 
\left({u \over {\cal R}} \right) 
\left( {d_p^{1/(7-p)} \over c_p^{1/(p-3)}} \right) \right]^{(p-3)(7-p) 
\over 2(5-p)}\, .
\end{equation}
We see that by making $N$ large, we can always make the dilaton small,
except when we approach the points $u=0$ or $u=\infty$, depending on $p$.
This analysis coincides with \cite{BST} and we refer to
that paper for a more detailed discussion of all D$p$--branes with
$p<8$. We have summarized some of the results in Table 6.

We would now like to focus our attention on the only D$p$--brane 
missing in previous discussions, the D8--brane. 

We find for the D8-dilaton background
\begin{equation}
e^{\phi}= {1\over N}\, \left[g_f^2
\left({u \over {\cal R}}\right)^{5}\left({16 \over \pi}\right)\right]^{1/6}\, 
\end{equation}
which means that the the dilaton background is small everywhere as long
as we take large $N$. This is necessary anyway since, in order 
to keep $g_f^2$ for the 
D8--brane fixed, we need $N \rightarrow \infty$. This means that we consider 
the 't Hooft limit of the (non-renormalizable) field theory.
Only when we approach $u=\infty$ the dilaton grows and we have to 
consider the S-dual situation which would involve M9--branes in M-theory.
We would like to point out that in the limit $N \rightarrow \infty$, the
size of the region where the dilaton is large can be made as small as
a point.
In the present case of the D8--brane
$u=\infty$ is the actual position of the D8-domain
wall, as opposed to branes with $p<6$, where $u=0$ denotes the position
of the brane.

For the dimensionless effective coupling (\ref{perturb}) and the 
string tension (\ref{sugravalid}) we find 
\begin{equation}
g^2_{\rm eff} = g^2_f u^5, \quad \tau_{D} = 
\left({1 \over 4 \pi} \left[g_f^2 \left({u \over {\cal R}}\right)^{5}
\left({16 \over \pi}\right)\right]^{1/6} \right)^{-2}, 
\end{equation}
so that supergravity and perturbative Yang-Mills are valid in the
same region for the D8--brane, near $u=0$:
\begin{eqnarray}
\label{D8valid}
u^5 \ll 1/g_f^2 &\rightarrow& \qquad FT \, , \nonumber \\
u^5 \ll 1/g_f^2 &\rightarrow& \qquad SUGRA \, .
\end{eqnarray}
This also happens for the D6--brane (although in that case we need to 
go to the strong coupling background which involves the $D=11$ KK--monopole
and in $D=11$ gravity does not decouple) 
and would suggest a contradiction in the sense that we have
two, apparently different, weakly coupled descriptions.
One reason could be that the decoupling of gravity of the field theory in 
these cases is rather subtle \cite{IMSY, BST}.  
However as we discussed earlier, when fixing the 't Hooft coupling 
constant, the decoupling of gravity in the 
{\it 't Hooft limit} of the field theory seems guaranteed for all branes.
One reason not to trust the decoupling of gravity is the 
fact that $g_s N$ is the open string coupling constant, which becomes
infinite in this case. Two open strings can join to form closed strings
(which includes gravitons). It is hard to understand why the strongly
coupled open string theory would not produce these closed strings 
(and thus decouple from gravity) in the low energy limit\footnote{
We thank K. Skenderis for a discussion on this point.}.
For now however we conjecture that
$DW_{10}$ supergravity near $u=0$ 
{\it equals} $D=9$ perturbative 
SYM theory {\it in the 't Hooft limit}. 
Another proposal for the interpretation of this result, 
although not unrelated to the `decoupling of gravity' argument, can be 
found in \cite{pepo}.
In any case, it is interesting to note
that the D6- and D8--brane behave in similar ways, a point of view that
is also advocated from considering the embedding of these branes in M-theory 
(where they both develop an extra {\em compact} direction)
\cite{on9branes}.

We end this subsection with Table 6 in which we summarise the 
results for ten-dimensional D$p$--branes.
{\small
\begin{center}
\begin{tabular}{|c|c|c|c|c|c|}
\hline
D$p$ & 0,1,2&3&4&6&8\\
\hline
$g_{eff}\ll 1$&$\infty$&$\forall$&0&0&0\\
$\tau_D \gg 1$&0&$\forall$&$\infty$&0&0\\
$e^\phi \gg 1$&0& $\emptyset$&$\infty$&0&$\infty$\\
$N $&$\forall$ & $\gg 1$ & $\infty$ &$\infty$ &$\infty$\\
\hline
\end{tabular}
\end{center}
\noindent {\bf Table 6}: The Table shows for which values of $u$, 
in the near-horizon geometry,
the restrictions I, II and III of Subsection 3.1,
i.e. $e^\phi \ll 1, \tau_D \gg 1$ and $g_{\rm eff} \gg 1$, are satisfied
for the case of D$p$--branes in $D=10$ dimensions.
The bottom row indicates the behaviour of $N$ (see also the discussion in 
the text).
\bigskip
}

\subsection{Six-dimensional d$p$--branes}  

Using the constraint $x=\Delta-\tilde d=p-1$, the relevant
dynamics should be governed by scalars.
Because these d$p$--branes are related to intersecting D$p$--branes the
worldvolume theory should again contain supersymmetric Yang-Mills. However,
the Yang-Mills coupling constant in this case is not kept fixed but
instead diverges. As we mentioned in Subsection 3.1, when we consider the
d$p$--branes on top of each other, we are in the Higgs branch vacuum 
of the theory. In the limit of diverging Yang-Mills coupling
constant all the vectors decouple and we are left with a scalar 
theory.
The (scalar) coupling constant
we keep fixed is 
\begin{equation}
g_f^2=c_p N g_s \sqrt{\alpha^\prime}^{\,p-1}\, .
\end{equation}
Other parameters we need are
\begin{equation}
\Delta=2 \, , \quad \tilde d = 3-p \, ,  \quad \beta= 2-p \, , 
\quad a= (1-p)\, .
\end{equation}

The quantities $u$ and $U$ can both be fixed at the same time.
In this specific case we obtain
\begin{equation}
U^{2-p} = g_f^2  \left({u \over {\cal R}}\right) \left({d_p \over c_p}\right)
\, .
\end{equation}
We see that until now the analysis resembles quite accurately that 
of the $D=10$ D$p$--branes.

In the following we will ignore the cases $p=1$ (no dilaton), $p=2$ 
(conformally flat, see Subsection 3.3) and $p=3$ ($\tilde d=0$).
To keep $g_f^2$ fixed for $p=0$ we find that it is possible but not necessary 
for $N$ to go to infinity. For $p=4$ we have to take $N$ to infinity 
to keep $g_f^2$ fixed.

Substituting the different parameters the dilaton background
(\ref{dilimit}) becomes
\begin{equation}
e^{\phi}= {1 \over N} \left[(g_f^2)^{1 \over p-1} 
\left({u \over {\cal R}} \right) \left( {d_p^{1/(3-p)} \over c_p^{1/(p-1)}} 
\right) \right]^{(p-1)(3-p) \over 2(2-p)}\, .
\end{equation}

Focusing on the d0-- and d4--brane, we next  analyse where the field
theory and supergravity description are valid using
(\ref{perturb}) and (\ref{sugravalid}). For the d0--brane we find that
the perturbative field theory description is valid 
near $u=\infty$ and the (S-dual) supergravity description is valid near 
$u=0$:
\begin{eqnarray}
\label{d0valid}
u \gg g_f^2 &\rightarrow& \qquad FT \, , \nonumber \\
u \ll g_f^2 &\rightarrow& \qquad SUGRA \, .
\end{eqnarray}
This is the typical behaviour in field theory--supergravity dualities,
both descriptions being valid in different regions, avoiding
contradictions. For the d0--brane it mimics the ten-dimensional
D0--brane situation.  As is clear from the above we should turn to the
S-dual background in the infrared. But what is the S-dual background?
We know that IIA superstring theory on a $K3$ manifold is S-dual to
Heterotic superstring theory on a $T^4$. The S-dual solution is
nothing but the 0--brane of Heterotic string theory on $T^4$, where the
extreme black hole is charged with respect to a Kaluza-Klein gauge
field or with respect to a $U(1)$ subgroup of the $SO(32)$ gauge
fields. This solution is fundamental in the sense that its mass is
independent of $g_s$ (like the fundamental string) and it has a
curvature singularity at $u=0$.  This curvature singularity should be
resolved by the `strong coupling limit' of the corresponding quantum
mechanics on the d0--branes. In this respect this situation resembles
the D1-F1 situation in IIB superstring theory in $D=10$
\cite{IMSY}. It would be interesting to explicitly determine the
infrared limit of the corresponding quantum mechanics model. We will
say more about this case in the next Section when we discuss 0--branes
in various dimensions.

For the `d4--brane' the situation is similar to that of the D8-domain-wall
in $D=10$. We find that
the field theory and supergravity description are valid in
the same region near $u=0$ 
apparently leading to a contradiction:
\begin{eqnarray}
\label{d4valid}
u^3 \ll 1/g_f^2  &\rightarrow& \qquad FT \, , \nonumber \\
u^3 \ll 1/g_f^2  &\rightarrow& \qquad SUGRA \, .
\end{eqnarray}
As was explained in the previous subsection, we have no good 
explanation for this puzzling situation. Our results suggest
the equivalence between $DW_{6}$ supergravity near $u=0$ 
and a $D=5$ perturbative (scalar) field theory, which is the Higgs branch of 
a $D=5$ SYM theory with 8 supercharges, in the 't Hooft 
limit ($N \rightarrow \infty$).
For arguments in favor of another interpretation of this result 
we refer to \cite{pepo}. 
Near $u=\infty$ we should use the S-dual background, which should be
embedded in Heterotic string theory on $T^4$. This involves the $D=10$
Heterotic solitonic 5--brane with one leg of the 5--brane wrapped around 
the $T^4$. Note that the $D=10$
interpretation of the d4--brane is a delocalized D4-D8 brane solution of Type
$IIA$ string theory. Duality conjectures based upon
a localized D4-D8--brane system embedded in Type $I^\prime$ string theory
have been discussed in \cite{AdS61,AdS62}.

This finishes our discussion on d$p$--branes preserving 8
supersymmetries. All features
of $D=10$ D$p$--branes reappear in the $D=6$ context. 
In Table 7 we have summarised our results on the six-dimensional 
`d$p$--branes'.

{\small
\begin{center}
\begin{tabular}{|c|c|c|c|}
\hline
d$p$ & 0& 1 &4\\
\hline
$g_{eff} \ll 1$&$\infty$&$\forall$&0\\
$\tau_D \gg 1$&0&$\forall$&0\\
$e^\phi \gg 1$&0&$\emptyset$&$\infty$\\
$N$&$\forall$&$\gg 1$&$\infty$\\
\hline
\end{tabular}
\end{center}
\noindent {\bf Table 7}: The Table shows for which values of $u$, 
in the near-horizon geometry,
the restrictions I, II and III of Subsection 3.1,
i.e. $e^\phi \ll 1, \tau_D \gg 1$ and $g_{\rm eff} \gg 1$, are satisfied
for the case of `d$p$--branes' in $D=6$ dimensions.
The bottom row indicates the behaviour of $N$ (see also the discussion in 
the text).
\bigskip
}

\section{Quantum mechanics and 0--branes}

In this Section we will consider the special case of 0--branes in 
various dimensions.
In particular, we will 
construct the Hamiltonian of the (boundary) quantum mechanics.
We expect the Hamiltonian to
be invariant under (generalized) conformal transformations
because of the (conjectured) duality between the supergravity theory and the 
boundary theory. Because the metric is pure AdS, the isometries of the
metric will induce the conformal invariance in the boundary QM.
It is clear that a non-trivial dilaton breaks the conformal isometries
of the background. As will be explained we can only expect to find 
generalized conformal invariance \cite{jeyo} in those cases.

Models of (super--) conformal quantum mechanics 
have been studied in \cite{aff,fura}. 
The relation of these models with black holes was pointed out recently
in \cite{derix}. The exact map between
$D=2$ (pure) $AdS$ (quantum) 
supergravity and the quantum mechanics model at the boundary, requires
a special discussion and is not yet
well understood \cite{review,mms,sstrom,pkto}.  

We will first give a general discussion, applicable to any dimension.
Next we will focus on the special
supersymmetric cases in $D=10$ (the
D0--brane), $D=6$ (the d0--brane, already partially discussed in the 
previous Section) and the pure AdS cases in $D=5$ (the m0--brane) and
$D=4$ (the Reissner-Nordstrom extreme black hole).  The RN-black hole
and the D0--brane have already been discussed in 
\cite{derix} and \cite{youm}, respectively. We will show that
these cases fit naturally into our general analysis.

To construct the quantum mechanics Hamiltonian we consider
a charged probe particle in the (supersymmetric) near-horizon background of 
$N$ stacked 0--branes.
Whenever we have a nontrivial dilaton we will restrict ourselves to
those cases for which the conditions (\ref{sufcon}) which for $p=0$
read $k=1$ and $x=\Delta - D +3$.
The (bosonic) Lagrangian for a particle with mass $m$ and charge $q$, 
moving in the dual frame (near-horizon) background of $N$ stacked 0--branes
then reads

\begin{equation}
{\cal L}= m \,  L g_s^{-1 \over D-3} e^{-{D-4 \over D-3} \phi} 
\sqrt{| \dot{x}^{\mu} \dot{x}^{\nu} g_{\mu\nu}^{(D)} | } +q A_{\mu} 
\dot{x}^{\mu} \, , 
\end{equation}

where the dot represents derivatives with respect to the worldline time
and $L$ is the (dimensionless) parameter defined in Appendix A (see 
also Section 3). The D-dimensional 0--brane solution in the
dual frame metric $g_{\mu \nu}^{(D)}$ is given by the expression
(\ref{limitsol}), taken for $p=0$. This expression
is constant in $\alpha^\prime$ units.
Introducing the canonical momentum $P_{\mu}={\partial {\cal L} \over  
\partial \dot{x}^{\mu}}$ we can write down the mass-shell constraint
for the probe particle which is

\begin{equation}
\label{mass-shell}
(P_{\mu} -qA_{\mu})(P_{\nu} -qA_{\nu})\, g^{\mu \nu}_{D} =
m^2 \, L^2 g_s^{-2 \over D-3}  e^{-2{ D-4 \over D-3}\phi} \, .
\end{equation}

We would like to solve this equation for $P_t=-{\cal H}$.
Substituting the background metric (\ref{limitsol}) into the mass shell 
equation (\ref{mass-shell}) and using the fact that in the static gauge
$A_\mu {\dot x}^\mu = A_t$, we solve for $P_t$ and find the Hamiltonian

\begin{equation}
{\cal H}={u \over {\cal R}} \sqrt{ \left( {u \over {\cal R}} \right)^2
{P_u}^2 + \vec{L}^2 + \alpha^\prime\, m^2 \, L^2 g_s^{-2 \over D-3}  
e^{-2 {D-4 \over D-3} \phi} } -qA_t \, ,
\end{equation}

where $\vec{L}^2 \equiv P_i P_j g^{ij} \alpha^\prime$, the (dimensionless) 
squared angular momentum vector over the sphere, not to be confused with $L$ 
which is just a constant.
Using the identity $A-B={A^2 -B^2 \over A+B}$
we can write this Hamiltonian as 

\begin{equation}
\label{conham}
{\cal H}={{P_u}^2 \over 2f}+{g \over 2f} \, ,
\end{equation}

with $f$ and $g$ given by 

\begin{eqnarray}
\label{fg}
f &=& \half e^{-{D-4 \over D-3} \phi} \left( {{\cal R} \over u} 
\right)^3 \left\{ \sqrt{\alpha^\prime m^2  (d_p N)^{2 \over (D-3)} 
+ e^{2{D-4 \over D-3} \phi} \left( {u \over {\cal R}} \right)^2 
\left[ P_u^2 +{\cal R}^2 {\vec{L}^2 \over u^2} \right] } \right. \nonumber \\
&& \hspace{4truecm} \left. + \left( {{\cal R} \over u} \right) 
e^{{D-4 \over D-3} \phi} A_t q \right\}\, , \\
g &=& e^{-2{D-4 \over D-3} \phi} \left( {{\cal R} \over u} \right)^2 
\left\{ \alpha^\prime m^2 (d_p N)^{2 \over D-3} - \left( {{\cal R} 
\over u} \right)^2 e^{2{D-4 \over D-3} \phi} {A_t}^2 q^2 + 
e^{2{D-4 \over D-3} \phi} \vec{L}^2 \right\} \, . \nonumber
\end{eqnarray}

We would like to substitute the expressions for the different background
fields and see if we can simplify the expressions for $f$ and $g$.
Note that we use the electric potential instead of the
dual magnetic potential we are using in the rest of the paper.
The solution for the (electric) vector potential is  

\begin{equation}
A_t= (g_s H)^{-1} \, .
\end{equation}

The expression for the dilaton is given in (\ref{limitsol}), but
we restrict ourselves to cases with $k=1$ and $x=\Delta - D +3$ so
we can use the dilaton expression in (\ref{dilback}). 
The reader may now check that 
for all supersymmetric 0--branes in the near-horizon limit, 
the following identity holds:

\begin{equation}
{{\cal R} \over u} A_t \,e^{{D-4 \over D-3} \phi} =
(d_p N)^{1/(D-3)}\,\sqrt{\alpha^\prime}\, ,
\end{equation}

This identity enables us to simplify the expressions 
for $f$ and $g$ considerably. Introducing a function ${\cal V}(u)$ which is 
equal to 

\begin{equation}
\label{vf}
{\cal V}(u)=(d_p N)^{2 \over D-3}\, e^{-2 {D-4 \over D-3} \phi} 
\left( {{\cal R} \over u} \right)^2 \, ,
\end{equation}

we can write the functions $f$ and $g$ as follows:

\begin{eqnarray}
\label{fgv}
f &=& \half \sqrt{{\cal V}(u)}\, \left( {{\cal R} \over u} \right)^2
\left\{ \sqrt{\alpha^\prime m^2 + {1\over {\cal V}(u)} \left[ P_u^2 +{\cal R}^2
{\vec{L}^2 \over u^2} \right] } +q \sqrt{\alpha^\prime} \right\}\, , 
\nonumber \\
g &=& {\cal V}(u)\, \alpha^\prime (m^2 -q^2) + {\cal R}^2 {\vec{L}^2 
\over u^2} \, .
\end{eqnarray}

From this form of $f$ and $g$ it is clear that in the field theory
limit ($\alpha^\prime \rightarrow 0$) the quantum mechanics model is
well defined because $m$ and $q$ have (implicit) $\alpha^\prime$ dependence
($m,q \sim 1/\sqrt{\alpha^\prime}$) 
cancelling all $\alpha^\prime$'s appearing in $f$ and $g$.
We also see that in the extreme limit ($m^2=q^2$) the function
$g$ simplifies considerably: $ g = {\cal R}^2 {\vec{L}^2 
\over u^2}$. This is the centrifugal potential for a free particle.
Except for a `conformal' factor in $f$ (depending on $u$) and the somewhat
unusual $u$ dependence of ${\cal V}(u)$, this should be a Hamiltonian 
of (`relativistic') conformal quantum mechanics as described in 
\cite{derix} as long as we keep the dilaton term in (\ref{vf}) fixed 
under conformal transformations. The appearance
of the $u$--dependent `conformal' factor (and the particular 
$u$--dependence of ${\cal V}(u)$) is the result of our choice of phase space
canonical variables $(u, P_u)$ and can be changed by making another
choice of canonical variables. Because ${\cal V}(u)$ is
just a power of $u$, when we take $u' \propto{\cal V}(u)^{1/4}$ and rewrite 
the Hamiltonian using the canonical variables $u'$ and $P_{u'}$, we 
remove the position dependence in this `conformal' factor in $f$.
The `non-relativistic' limit \cite{derix} is one in which the function
${\cal V}(u)$ goes to infinity and the expression ${\cal V}(u)\,
\alpha^\prime (m^2 -q^2)$ is kept fixed. In that limit, except for 
an infinite `conformal' factor which can be removed by using another
canonical set of variables, we obtain the conformal quantum mechanics
of \cite{aff}. 

The generators of special conformal and dilatation transformations
are equal to

\begin{equation}
{\cal K}= f u^2 \quad , \quad {\cal D}= u P_u \, .
\end{equation}

Together with the generator of time translations, the Hamiltonian
${\cal H}$, these generators satisfy the following $SL(2,R)$ algebra
\cite{fura}

\begin{equation}
\label{sl2r}
[{\cal H},{\cal D}]={\cal H}\,, \quad [{\cal D}, {\cal K}]={\cal K}\, , \quad
[{\cal H}, {\cal K}]= {\cal D} \, .
\end{equation}
This is only true when the effective coupling constant $g_{eff}^2$
of the QM model under consideration is fixed under scale transformations. 
Whenever we have a non-trivial 
dilaton the QM model will not be conformal invariant by itself. 
Only when introducing a transformation of $g_f^2$ to keep $g_{eff}^2$ 
fixed under conformal transformations, will the QM model
be invariant under what are called generalized conformal transformations
\cite{jeyo}.

We will now discuss as special cases the
supersymmetric black holes in $D=10, 6, 5, 4$. 

\begin{itemize}
\item{ {\it The $D=10$ D0--brane}. \hskip .3truecm
For $D=10$ the function ${\cal V}(u)$ is
given by

\begin{equation}
{\cal V}(u)= (N d_p)^{\frac{4}{5}} \left( {N c_p \over {\cal R}^3 g_{eff}^2} 
\right)^{\frac{6}{5}} \left( {{\cal R} \over u} \right)^2 \, .
\end{equation}

Remember that there is some $u$ dependence hidden in $g_{eff}^2$ and
that the model is invariant under generalized conformal transformations
keeping $g_{eff}^2$ fixed.
The expression differs from the one given in 
\cite{youm}. Performing a 
coordinate transformation, going from the variables $u$ and $P_u$ to the 
canonical variables $U$ and $P_U$ (using (\ref{uU})) we loose the position 
dependence of the `conformal' factor in $f$ and we obtain the result 
of \cite{youm}. We refer to that paper and \cite{pkto}
for more details on the relation 
of this model with (M)atrix theory.}
\item{ {\it The $D=6$ d0--brane}. \hskip .3truecm
For $D=6$ the function ${\cal V}(u)$ is given by

\begin{equation}
{\cal V}(u)= N d_p \left( {N c_p \over {\cal R} g_{eff}^2} \right) 
\left( {{\cal R} \over u} \right)^2 \, .
\end{equation}

As in the previous example, there is hidden $u$ dependence in $g_{eff}^2$
and the model is therefore invariant under generalized conformal 
transformations keeping $g_{eff}^2$ fixed.
The canonical set of variables $(r, P_r)$ will remove the position 
dependence of the `conformal' factor in $f$ if $u \propto r^{-4}$ (this 
means we can not use $U$, which is proportional to $\sqrt{u}$, as we did
in the D0--brane case). Considering the extremal case and purely radial
motion ($\vec{L}^2=0$), the large $u$ (small ${\cal V}$) limit should 
be considered `ultra-relativistic' and the small $u$ (large ${\cal V}$) 
limit should be considered a `classical' limit of the quantum mechanics 
model in the sense that the correction terms to $m$ in the function $f$
can be neglected. This suggests that the singularity in the Heterotic
0--brane solution (see Subsection 4.2) is resolved by a (`classical') 
free conformal quantum mechanics model. This would be
in correspondence with the 
resolution of the singularity in the fundamental string solution by a free 
orbifold conformal field theory \cite{dvv}.
}
\item{ {\it The $D=5$ m0--brane}.  \hskip .3truecm
For $D=5$ the function ${\cal V}(u)$ is
given by

\begin{equation}
{\cal V}(u)= N d_p \left( {{\cal R} \over u} \right)^2 \, .
\end{equation}

The main difference with the previous cases is the fact that
we no longer have hidden $u$ dependence in some effective
coupling constant $g_{eff}^2$. This QM model is conformal invariant without 
having to introduce an extra scaling law for $g_f^2$ which is
a consequence of the fact that the background is pure AdS. }

\item{ {\it The $D=4$ extreme RN-black hole}.  \hskip .3truecm
For $D=4$ the function ${\cal V}(u)$ is

given by

\begin{equation}
{\cal V}(u)= (N d_p)^2 \left( {{\cal R} \over u} \right)^2 \, ,
\end{equation}
which is the same as the expression for the m0--brane, except for 
a factor of $N d_p $. Again this is not the same as the expression
in \cite{derix}, because of the different phase space variables we
are using. }
\end{itemize}

So far, we have been considering the conformal quantum mechanics model
of a single 0--brane. It has been argued recently \cite{GTCM} that the
quantum mechanical model governing the fluctuations of $N$ stacked
0--branes is given by the $N$--particle Calogero model.
This is based on the observation that the $N$--particle
Calogero model is related, in the large N limit, to the reduction of
a $D=2$ Yang-Mills theory describing the fluctuations of N stacked 1--branes.

The expression of the $N$--particle Calogero Hamiltonian, acting
on the
subspace of totally symmetric wavefunctions, is given by

\begin{equation}
{\cal H} = {\textstyle {1\over 2}}\sum_i p_i^2 + \sum_{i<j} {\ell(\ell-1)\over 
(q_i - q_j)^2}\, .
\end{equation}

Here $\ell$ is a constant and 
$(q_i,p_i)\ (i=1,\cdots N)$ are a set of 2N canonical variables.
In terms of the `coupled' momentum operators

\begin{equation}
\pi_i = p_i + \ell\sum_{i\ne j} {K_{ij}\over q_i - q_j},
\end{equation}
where $K_{ij}$ are the so-called exchange operators, 
the Calogero Hamiltonian can
be brought into the free form \cite{Pol}

\begin{equation}
{\cal H} = \sum_i \pi_i^2\, .
\end{equation}

In terms of annihilation  and creation operators $(a_i,a_i^\dagger)$
satisfying a so-called
$S_N$--extended Heisenberg algebra
the Hamiltonian can be written as \cite{BHV}

\begin{equation}
{\cal H} = {\textstyle{1\over 2}}\sum_i \{a_i, a_i^\dagger\}\,.
\end{equation}

In terms of these operators the generators of the $SL(2,R)$ algebra
(\ref{sl2r}), which is called the `vertical' $SL(2,R)$ algebra,
are given by the Calogero Hamiltonian ${\cal H}$ and the two operators

\begin{equation}
B_2^+ = {\textstyle{1\over 2}}\sum_i (a_i^\dagger)^2\, ,\hskip 2truecm
B_2^- = {\textstyle{1\over 2}}\sum_i (a_i)^2\, .
\end{equation}

The point we want to make here is that the Calogero Hamiltonian is also
part of another so-called `horizontal' $SL(2,R)$ algebra with generators
$\{L_{-1}, L_0, L_1\}$ such that $L_0 - {\cal H}$ is a pure constant
and the other two generators are given by

\begin{eqnarray}
L_1 &=& \sum_i a_i\, ,\cr
&& \cr
L_{-1} &=& \sum_i \left [ \alpha (a_i^\dagger)^2 a_i + (1-\alpha) a_i
(a_i^\dagger)^2\right ] + 2(\lambda - {\textstyle{1\over 2}})
\sum_ia_i^\dagger\, , 
\end{eqnarray}
with $\alpha,\lambda$ arbitrary constants. It turns out that this
`horizontal' $SL(2,R)$ algebra can naturally be extended to a 
Virasoro algebra which acts as a kind of spectrum--generating algebra
of the Calogero model. For more details, see \cite{BV}\footnote{
Recently, it has been shown that any scale-invariant mechanics
of one variable has the symmetries of a full Virasoro algebra
\cite{JK}.}. It would be
interesting to see whether this hidden Virasoro symmetry
will play a role in the microscopic description of black holes in terms of
the Calogero model.

%%%%%%%%%%%%%%%%%%%%%%%%%%%%%%%%%%%%%%%%%%%%%%%%

\section{The Type I/Heterotic 6--brane}

%%%%%%%%%%%%%%%%%%%%%%%%%%%%%%%%%%%%%%%%%%%%%%%%

In this Section we will discuss some properties of the Type I/Heterotic 
6--brane solution we discussed in Subsection 2.3 as an example
of a brane whose near-horizon geometry is a conformally flat space.
We will only discuss some classical aspects of the
solution and not the field theory limit so that this Section
can be read independent from the previous Sections.

The Type I/Heterotic 6--brane is magnetically charged under an U(1) subgroup
of the 10-d Yang-Mills gauge field. The complete solution reads
\be800
\ba{l}
ds^2_6 = -dt^2 + dy_1^2 + \cdots + dy_6^2 + H^2 \, d\vec x d\vec x\, ,\\ 
F_{mn} = \epsilon_{mnp} \partial_p H  \qquad , \qquad e^{2 \phi} = H\, ,
\ea
\ee
with $m,n = 1, 2, 3$ and $H$ is a harmonic function in the transverse
coordinates $\vec x = (x_1 , x_2 , x_3)$. The first thing to notice
is, that this solution is not supersymmetric in 10 dimensions.
Namely, the heterotic susy variations of the gravitino $\psi_M$, the
dilatino $\lambda$ and the gaugino $\chi$ are
\be810
\ba{rcl}
\delta \psi_{M} &=& \Big[ \partial_{M} + {1 \over 4} \Omega^{(-)\ AB}_{M}
\Gamma_{AB} \Big] \epsilon\, , \\
\delta \lambda &=& \Big[ - \Gamma^M \partial_M \phi + {1 \over 2 \, 3!}
\Gamma^{MNP} H_{MNP} \Big] \epsilon\, , \\
\delta \chi &=& \Gamma^{MN} F_{MN} \epsilon\, ,
\ea
\ee
where $\Omega^{(-)\ AB}_{M}$ is defined as a combination of the spin connection
and the torsion ($A,B$ are flat indices):

\begin{equation}
\Omega^{(-)\ AB}_{M} \equiv \omega_M^{AB} - H_M^{\ AB}\, .
\end{equation}
Since the torsion is trivial for the
Type I/Heterotic 6--brane solution, none of these variations can vanish (assuming that
the Killing spinor does not depend on the world volume directions).
So how is this solution related to the other {\em supersymmetric}
branes? In heterotic string theory we have only the NS5--brane and the
KK--monopole which give rise to magnetic charges and the bound state
of both is given by
\be820
\ba{l}
ds^2_{5\times KK} = -dt^2 + dy_1^2 + \cdots + dy_5^2 + H_1 \, 
\Big[ {1 \over H_2} (dx_4 + V_m dx^m)^2 + H_2 d\vec x d \vec x \Big]\, , \\
H_{4mn} = \epsilon_{mnp} \partial_p H_1 \qquad , \qquad
\partial_m H_2 = \epsilon_{mnp} \partial_n V_p
\qquad , \qquad e^{2 \phi} = H_1 \ .
\ea
\ee
Obviously, in order to obtain the 6--brane we must
identify the two harmonics: $H_1 = H_2$. However, upon reduction
to 9 dimensions the gauge fields coming from this bound state enter
only the right--moving sector (see e.g.\ \cite{810}), but the gauge field
coming from the 6--brane is part of the left--moving sector. Therefore,
there cannot be any heterotic O(2,1)-duality that relates both to each
other. Instead, in order to find a correspondence we have to make sure
that the 9-dimensional gauge fields are in the same sector, which can
be done by changing the sign of the torsion, i.e.\ we replace
(\ref{820}) by
\be830
\ba{l}
ds^2_{\bar 5 \times KK} = -dt^2 + dy_1^2 + \cdots + dy_5^2 +  
\Big[ (dx_4 + V_m dx^m)^2 + H^2 d\vec x d \vec x \Big]\, , \\
H_{4mn} = - \epsilon_{mnp} \partial_p H \qquad , \qquad
\partial_m H = \epsilon_{mnp} \partial_n V_p
\qquad , \qquad e^{2 \phi} = H\, , 
\ea
\ee
where the subscript $\bar 5 \times KK$ indicates that we consider a
bound state of an anti-5--brane and a KK--monopole.  The equations of
motions are invariant under this sign change, but it breaks
supersymmetry! After reducing over $x_4$ the
9-dimensional gauge fields have now the same chirality and can be
rotated into each other. In fact the 6--brane is related to (\ref{830})
by the $O(2,1)$ transformation
\renewcommand{\arraystretch}{1.2}
\be840
e^{- {1 \over \sqrt 2} J_-} e^{\sqrt 2 J_+}
=  \left(\ba{ccc} 1 & 1 & -\sqrt 2  \\ 1 & 1 & - 1/\sqrt 2 \\
1/\sqrt 2 & - 1/\sqrt 2 & 0 \ea \right)\, ,
\ee
where $J_{\pm}$ are the boost generators of $O(2,1)$ (we use
the notation of \cite{820}) and this transformation
relates the 9-dimensional gauge fields to each other
\be850
\left(\ba{c} V_{\mu} \\ -V_{\mu} \\ 0 \ea \right)_{\bar 5 \times KK}
\  \leftrightarrow \  
\left(\ba{c} 0\\0 \\ {\sqrt 2} \, V_{\mu} \ea \right)_6\, .
\ee
\renewcommand{\arraystretch}{1.7}
There is no Wilson line created by this duality, i.e.\ $V_x = 0$.
The dilaton and the compactification 
radius ($R^2 = g_{44}$) are related to each other by
\be860
e^{2 \phi_6} = {1 \over 2} \, e^{2 \phi_{\bar 5 \times KK}}
\qquad , \qquad R_6 = {1 \over 2} \, R_{\bar 5 \times KK} = {1 \over 2}\, ,
\ee
i.e.\ the string coupling constant as well as the compactification
radius are only half of the original values. 
Thus, the 6--brane can be seen as a bound state of an anti-5--brane and
a KK--monopole at the self-dual radius (defined by $H_1=H_2$ giving
$R_{\bar 5\times KK} =1$), as discussed in
\cite{830}. Related examples of non-supersymmetric brane solutions
have been dicussed in \cite{832,834}.

It is easy to construct intersections of 6--branes.
Since the worldvolume
directions trivially factorizes for the 6--brane, it is straightforward
to combine up to three 6--branes
\be870
\ba{l}
ds^2_{6\times 6 \times 6} = -dt^2 + H_1^2 d\vec x_1 d\vec x_1 +
 H_2^2 d\vec x_2 d\vec x_2 + H_3^2 d\vec x_3 d\vec x_3\, , \\
F_{mn}^{i} = \epsilon_{mnp} \partial_p H_i  
\qquad , \qquad e^{2 \phi} = H_1 H_2 H_3\, ,
\ea
\ee
where $\vec x_i$ are the coordinates of three 3-dimensional subspaces ($i =
1 \cdots 3$). We see, that two 6--branes intersect over a 3--brane
(common worldvolume) and all three intersect over a point.
Introducing polar coordinates the harmonic functions can be written as
\be880
H_i = 1 + {q_i \over r_i} \ .
\ee
Taking the limit where we can ignore the constant parts in 
$H_i$ and introducing $r_i = q_i e^{q_i y^i}$ the metric and dilaton
becomes
\be890
ds^2_{6\times 6 \times 6} = -dt^2 + dy_1^2 + dy^2_2 + dy_3^2 + 
q_i^2 d\Omega_2^{(i)} \qquad , \qquad e^{2 \phi} = e^{q_i y^i} \ .
\ee
Hence, we get a linear dilaton and the geometry is $M_4 \times S_2
\times S_2 \times S_2$ where $M_4$ is a flat 4-dimensional Minkowski
spacetime.

The Type I/Heterotic 6--brane has also an electric dual, which can be
constructed by reducing the 6--brane to 4 dimension and applying
the standard rules of S-duality and a subsequent oxidation to 10
dimensions. As a result one finds the following
Type I/Heterotic 0--brane solution:
\be900
\ba{l}
ds^2_0 = -{1 \over H^2} dt^2 + dy_1^2 + \cdots + dy_9^2\, , \\
A_0 = 1/H \qquad  , \qquad e^{-2\phi} = H \ .
\ea
\ee
In analogy to the 6--brane this Type I/Heterotic 0--brane can be
obtained via an $O(2,1)$ T-duality {from} an anti-string -- wave
bound state at the self-dual radius.  Using the $O(2,1)$
transformation it is straightforward to construct a bound state
of a 6- and a 0--brane which is given by
\be910
ds^2 = -{1 \over H_0^2} dt^2 + H^2_1 \, d\vec x d\vec x + dy_1^2 + \cdots
+ dy^2_6 \ .
\ee
This solution can immediately be verified by compactifying all
$y$ coordinates which yields the standard dyonic Reissner-Nordstr\"om
black hole.

%%%%%%%%%%%%%%%%%%%%%%%%%%%%%%%%%%%%%%%%%%%%%%%%%%%%%%%%

\section{Conclusions}

%%%%%%%%%%%%%%%%%%%%%%%%%%%%%%%%%%%%%%%%%%%%%%%%%%%%%%%

In this paper, we presented a systematic overview of the near-horizon
geometry of any two-block $p$--brane solution and discussed the corresponding
domain-wall/QFT duality, thereby generalizing the discussion of \cite{IMSY,BST}.
The near-horizon geometry of the general $p$--brane solution is singular, 
but the
singularity is only contained in a conformal factor; for all
$p$--branes the Weyl-tensor is regular.  In fact, in a conformally
rescaled so-called `dual' frame, the spacetime factorizes into an 
anti-de Sitter space
and a sphere \cite{9405124,BPS}. 
The analysis of appendix B shows, that this `dual frame' 
metric coincides with the sigma-model metric \cite{duff}, which enters the
Nambu-Goto action of a dual brane probe \cite{9405124}.
It is for this reason that the metric is called a `dual frame' metric.

By reducing over the spherical part of the $p$--brane near-horizon geometry
one obtains a domain-wall solution. For any $p$--brane
we established the explicit form of this domain-wall in terms of the
parameters of the original $p$--brane solution.
There are two special cases.  
First, the domain-wall can become an AdS space, which
corresponds to the well-known regular branes with a trivial dilaton.
Second, the domain-wall can become  flat. In 10 dimensions this is
the case for the 5--branes and for the Heterotic/Type I 6--brane. 
The latter case is new and was discussed in Section 6. In particular,
we showed that the Heterotic/Type I 6--brane is T-dual to the bound state of an
anti-5--brane and a KK monopole at the self-dual radius and thus it
is not supersymmetric.

The main goal of this work was to discuss the field theory limit 
of $p$--branes in a general setup. The 
starting point for the analysis was the fact that the background
should be regular and finite in $\alpha^\prime$ units in the low
energy limit. We argued that in order for the DW/QFT dualities to be well
behaved we need to restrict ourselves to $D=10$ D--brane intersections 
reduced over their relative directions (with identified harmonics).
Starting with that working assumption we added new
cases to the list of (conjectured) dualities between field theory
and supergravity (in various dimensions) on a 
${\rm DW}_{d+1} \times S_{\tilde d-1}$ background with a non-trivial
dilaton.

One new case we discussed is the domain-wall brane in $D=10$ and $D=6$
dimensions, i.e. the D8--brane and the d4--brane. A special
feature of domain-walls is that the `near-horizon' limit
$H \rightarrow \infty$ corresponds to the asymptotic limit
$r\rightarrow \infty$ instead of $r \rightarrow 0$ as for all the
other $p$--branes. We
found that the D8--brane is similar to the D6--brane in the sense that
the field theory and supergravity description are both valid
in the infrared limit $u=0$. However, an important difference is that 
for the D6--brane
we have $e^\phi \gg 1$ at $u=0$ and one must use the S-dual
KK--monopole description whose near-horizon geometry is the flat
Euclidean 4-space $\E^4$ without a boundary \cite{TL}. On the other
hand, for the D8--brane we have $e^\phi \ll 1$ at $u=0$ and one can
stay in ten dimensions. This leads to
a duality between $D=10$ (massive) supergravity and (the large
N limit of) $D=9$ Yang-Mills. This duality seems unlikely and
the D8--brane is at present not well understood.
Due to its relation with the M9--brane \cite{on9branes} this case
clearly deserves further study.

In this work we did not discuss the boundary field theories
in detail except for the extreme black holes, where we constructed
the (generalized conformal) quantum mechanics Hamiltonian. 
An interesting case is 
the d0--brane in $D=6$ dimensions. We find that in the infrared
limit one must use the S-dual fundamental Heterotic 0--brane description.
We argued that the curvature singularity of this fundamental 0--brane,
in correspondence with the curvature singularity of the
fundamental $D=10$ string, is resolved by a free (conformal) 
quantum mechanics model (see Section 5). It would be interesting
to study more precisely the mechanism behind this curvature resolution.

Following the conjecture of \cite{GTCM} we considered the $n$-particle
Calogero model as the conformal quantum mechanics model describing
$n$ stacked Black Holes. We pointed out that the `horizontal'
$SL(2,\R)$ symmetry of the Calogero model can be extended to a Virasoro
Algebra. It would be interesting to see whether this Virasoro symmetry
can lead to any new insight into the microscopic description of
black holes. 

The results of this paper have been obtained using a general setup. 
Our hope is that
the present work will be a convenient starting point 
for investigating in more detail the validity of the
different conjectured DW/QFT
dualities. Several issues involved in these conjectures have
not been discussed here. For instance, in some cases the worldvolume
QFT involved in the DW/QFT duality
is non-renormalizable. This indicates, for particular regions
in the parameter space, the presence of extra
non-decoupled degrees of freedom similar to the ones discussed in
\cite{hullim}. In this paper we discussed duality conjectures which 
can be related to {\it delocalized} D$p$--brane intersections. Due to this
we have not been very precise about the vacuum structure 
of the dual quantum field theory. To improve this one should in some cases 
consider {\it localized} intersections as discussed in 
\cite{AdS61,AdS62,mape,AdS4}.
We hope to report on this and other issues
in the near future.

%%%%%%%%%%%%%%%%%%%%%%%%%%%%%%%%%%%%%%%%%%%%%%%%%%%%%%%%%%%

\section*{Acknowledgements}

We wish to thank C.M. Hull, E. Sezgin, K. Skenderis, P. Sundell
and P.K. Townsend for useful discussions.
This work is supported by the European Commission TMR programme
ERBFMRX-CT96-0045, in which E.B., R.H. and J.P.v.d.S. are  associated 
to the University of
Utrecht. The work of R.H. and J.P.v.d.S. is
part of the research program of the ``Stichting voor Fundamenteel
Onderzoek der Materie'' (FOM).
J.P.v.d.S. wishes to thank the Physics Department of the
Humboldt University in Berlin,
where part of this work was done, for
its hospitality. K.B. wishes to thank the Physics Department of Groningen
for its hospitality.

\newpage

%%%%%%%%%%%%%%%%%%%%%%%%%%%%%%%%%%%%%%%%%%%%%%%%%%%%%%%%%%%%%%%%%%%%%%

%%%%%%%%%%%%%%%%%%%%%%%%%%%%%%%%%%%%%%%%%%%%%%%%%%%%%%%%%%%%%%%%%%%%%%
\appendix
%%%%%%%%%%%%%%%%%%%%%%%%%%%%%%%%%%%%%%%%%%%%%%%%%%%%%%%%%%%%%%%%%%%%%%

\section{Charge Conventions}

In this Appendix we specify our charge conventions.
In $D=10$ we use the convention 
$l_s^2 = \alpha^\prime$ which is understood to be replaced by the 
Planck length $l_p$ in other dimensions.

The charge of a $p$--brane is defined by
\begin{eqnarray}
\mu_{p} & = & {1 \over \sqrt{16 \pi G_{D}}}
\int_{S_{\tilde d+1}} F_{\tilde d+1} \nonumber \\
        & = & {\Omega_{\tilde d+1} Q \over \sqrt{16 \pi G_{D}}},
\end{eqnarray}
where  the angular volume of a sphere at transverse infinity of the
 brane is given by
\begin{equation}
\Omega_{\tilde d + 1} = {2 \pi^{{\tilde d+2 \over 2}} \over
\Gamma({\tilde d+2 \over 2})}.
\end{equation}
 
Expressing the BPS relation between charge and mass in terms
of the flux $\mu_{p}$ and the effective tension $\tau_{p}$ , we have for $N$
coincident branes
\begin{equation}
\mu_{p} = N \sqrt{16 \pi G_{D}} \, \tau_{p}.
\end{equation}
When we define a basic charge $q$ with a (natural) unit of $[l]^{-1}$ as
\begin{equation}
Nq \equiv {\Omega_{\tilde d+1} V_{p}\, Q \over 16 \pi G_D} \, ,
\end{equation}
where $V_p$ is the spatial volume of the $p$--brane, and use 
that $\tau_p=m/V_p$, then the BPS relation in terms of the
quantities $m$ and $q$ naturally reads $m=q$.
 
Using scaling arguments (see Appendix B) we find for 
$\tau_{p}$ in the string frame
\begin{equation}
\tau_{p} = {2\pi \over (2 \pi l_s)^d } \, g_{s}^{-k} ,
\end{equation}
with 

\begin{equation}
k={a \over 2} + {2 d \over D-2}\, .
\end{equation}
  
We have for Newton's constant in $D$ dimensions:
\begin{equation}
16 \pi G_{D} = {(2 \pi l_s)^{D-2} \over 2 \pi} \, g_{s}^2 .
\end{equation}
 
Using the above we can express
$Q$ in terms of $g_{s}$ and $l_s$
 \begin{equation}
\label{charge}
Q = {(2 \pi l_s)^{\tilde d} \over \Omega_{\tilde d+1}} 
N g_s^{2-k}\, .
\end{equation}
 
We can relate $Q$ to $r_0$ by
\begin{equation}
Q = {\tilde d} \, r_0^{\tilde d}\, .
\end{equation}     

Finally, it is convenient to define parameters $L$ and $d_p$ by:
\begin{eqnarray}
d_p &=& {(2 \pi)^{\tilde d} \over \tilde d \, \Omega_{\tilde d + 1}}\, , \\
\label{numfactor}
L &\equiv& {r_0 \over l_s} = (d_p N g_s^{2-k})^{1 \over \tilde d} \, .
\end{eqnarray}

%%%%%%%%%%%%%%%%%%%%%%%%%%%%%%%%%%%%%%%%%%%%%%%%%%%

\section{The sigma model metric}

%%%%%%%%%%%%%%%%%%%%%%%%%%%%%%%%%%%%%%%%%%%%%%%%%%%%%%

In this paper we have been working with two 
kinds of metric: the Einstein metric $g_E$
and the regular `dual frame' metric $g_D$. 
The Einstein metric is defined as the metric
frame in which there is no dilaton factor in front of the Einstein term,
like in (\ref{105}). The regular metric was defined in (\ref{160})
and is the metric frame in which the $p$--brane solution (\ref{2block})
has a regular near-horizon metric, like in (\ref{standard}).

In this section we consider a third metric frame: the `sigma model'
metric $g_\sigma$ \cite{duff}. Given a $\tilde d$-form gauge potential $C$, 
it is
defined as the metric in which the sigma model action containing the 
$\tilde d$-form potential $C$ has a dilaton-independent effective brane tension
$\tau_{\tilde d -1}$. 
To be more precise, consider the action

\begin{equation}
\label{tot}
S^{\rm total} = S_D + I_D\, ,
\end{equation}
where $S_D$ is the target space action given in (\ref{105}) and
$I_D$ is the following sigma model action

\begin{eqnarray}
\label{sm}
I_D &=& T_{\tilde d -1}\int {\rm d}^{\tilde d}\xi\ \left\{
\sqrt {{\rm det} \partial_i X^M\partial_j X^N
g_{\sigma, MN}}\right.\\
&&\left. + \varepsilon^{i_1\cdots i_{\tilde d}}\, \partial_{i_1}X^{M_1}\cdots 
\partial_{i_{\tilde d}}X^{M_{\tilde d}}C_{M_1\cdots M_{\tilde d}}
 + \cdots\right\}\, ,\nonumber
\end{eqnarray}
with $T_{\tilde d -1}$ the brane tension parameter.
We have only indicated the standard Nambu-Goto term and the leading 
term of the Wess-Zumino contribution. The effective brane tension 
$\tau$ in the Einstein frame is given by

\begin{equation}
\tau_{\tilde d -1} = T_{\tilde d -1} \left [ \Omega(\phi) \right ]^{
\tilde d/2}
\end{equation}
and is independent of the dilaton only in the sigma model metric $g_\sigma$
given by 

\begin{equation}
g_\sigma = \Omega(\phi) g_E\, .
\end{equation}
To determine the scale factor $\Omega(\phi)$ we follow the procedure 
of \cite{duff} and require that the total action (\ref{tot}) scales 
homogeneously under the scaling transformations

\begin{equation}
g_E \rightarrow \alpha g_E\, ,\qquad
e^\phi \rightarrow \beta e^\phi\, ,\qquad
C_{\tilde d} \rightarrow \lambda^{\tilde d} C_{\tilde d}\, .
\end{equation}
Requiring that each term in the action scales with the same factor
we find:

\begin{equation}
\Omega \rightarrow {\lambda^2\over \alpha}\Omega\, ,\qquad
\alpha^{D/2-1} = \lambda^{\tilde d} = \alpha^{D/2-1-\tilde d}
\beta^{-a}\lambda^{2\tilde d}\, ,
\end{equation}
from which we deduce that

\begin{equation}
\Omega = e^{\left ({a\over \tilde d}\right )\phi}\, ,\qquad
\alpha = \lambda^{{2\tilde d\over D-2}}\, ,\qquad
\beta = \lambda^{{2d\tilde d\over (D-2)a}}\, .
\end{equation}
We thus find, in particular, that

\begin{equation}
g_\sigma = e^{\left ({a\over \tilde d}\right )\phi}g_E\, .
\end{equation}
Comparing with the definition (\ref{160}) of the regular `dual frame' metric
we conclude that

\begin{equation}
\label{id}
g_\sigma = g_D\, .
\end{equation}
In terms of $g_\sigma$ the precise scaling transformations are given by

\begin{eqnarray}
g_\sigma &\rightarrow& \lambda^2 g_\sigma\, ,\nonumber\\
e^\phi &\rightarrow& \lambda^{{2d\tilde d\over (D-2)a}} e^\phi\, ,\\
C_{\tilde d} &\rightarrow& \lambda^{\tilde d} C_{\tilde d}\, ,\nonumber
\end{eqnarray}
such that

\begin{equation}
S^{\rm total} \rightarrow \lambda^{\tilde d} S^{\rm total}\, .
\end{equation}

Note that the sigma model metric $g_\sigma$ is defined independent
of any solution but that the definition of the regular `dual frame'
metric $g_D$
is related to a particular solution. In the identity (\ref{id})
it is understood that $g_D$ is related to a $p$--brane solution
(in the magnetic formulation) whereas $g_\sigma$ refers to
a sigma model referring to the dual $(\tilde d -1)$--brane.

We conclude that a brane moving in a background of its dual brane does not
see any spacetime singularity. For instance,  a D2--brane in 10 dimensions
that
probes a (dual) D4--brane will enter an anti-de Sitter space in the
near-horizon region.  On the other hand probing a D4--brane with a
(non--dual) D1-string, that couples to the string metric, 
leads to a singular
geometry.  Only for the selfdual D3--brane all frames
are equivalent and one always sees a regular geometry. Only the D3--brane
has a zero dilaton, for all other D$p$--branes the dilaton is
singular indicating a strongly or weakly coupled region corresponding
to a free field theory either in the UV or IR fixed points.

We finally note that there is yet a third kind of metric: the usual
string frame metric $g_S$. It is defined as the
metric in which the Einstein term is
multiplied by a dilaton factor $e^{-2\phi}$ and the dilaton kinetic
term carries a factor $4$ in front. In terms of the
Einstein metric $g_E$ it is given by\footnote{The used Einstein frame
(\ref{105}) ensures that we obtain the string frame, after the 
conformal transformation as defined in (\ref{sframe}), with the 
correct factor $4$ in front of the dilaton kinetic term in all dimensions.}

\begin{equation}
\label{sframe}
g_S = e^{{4\over D-2}\phi}g_E\, .
\end{equation}
The string metric $g_S$ coincides with the sigma model metric
$g_\sigma$ and/or the regular metric $g_D$ if

\begin{equation}
{4\over D-2} = {a\over \tilde d}\, ,
\end{equation}
which includes for example the NS-5--brane for which $D=10, a = 1, 
\tilde d = 2$.

%%%%%%%%%%%%%%%%%%%%%%%%%%%%%%%%%%%%%%%%%%%%%%%

\section{Glossary of used symbols}

%%%%%%%%%%%%%%%%%%%%%%%%%%%%%%%%%%%%%%%%%%%%%

Due to our general setup this paper contains many symbols.
For the convenience of the reader we give in this Appendix
an overview of (most of) the symbols introduced in the different Sections.
Standard and/or obvious symbols have been omitted. Next to a short
description we have given the equation number in which the symbol
first occurred.
 
\begin{itemize}
\item {\bf Section 1}
{\small
\begin{center}
\begin{tabular}{|c|c|c|}
\hline
{\bf Symbol}& {\bf Short description} & {\bf Equation} \\
\hline
$d$& Dimension of $p$--brane worldvolume & (1)\cr
$b$& Domain-wall dilaton coupling parameter & (1)\cr
$\epsilon$& Free parameter of harmonic solution & (3) \cr
$\Delta_{DW}$& Domain-wall parameter invariant under reductions& (4)\cr
$\lambda$& Radial coordinate & (10)\cr
\hline
\end{tabular}
\end{center}
\bigskip
}
\newpage
\item {\bf Section 2}
{\small
\begin{center}
\begin{tabular}{|c|c|c|}
\hline
{\bf Symbol}& {\bf Short description} & {\bf Equation} \\
\hline
$a$& $p$--brane dilaton coupling parameter & (14)\cr
$\tilde d$& Dimension of dual $\tilde d-1$--brane worldvolume & (15)\cr
$\Delta$& $p$--brane parameter invariant under reductions& (17)\cr
$\delta$& Dual frame dilaton parameter & (24)\cr
$\gamma$& Dual frame dilaton kinetic parameter & (24)\cr
$u$& $AdS$ radial coordinate and holographic energy scale & (29)\cr
${\cal R}$& Dimensionless radius of $AdS$ & (29)\cr
$\beta$& Inverse of ${\cal R}$ & (29)\cr
\hline
\end{tabular}
\end{center}
\bigskip
}
\item {\bf Section 3}
{\small
\begin{center}
\begin{tabular}{|c|c|c|}
\hline
{\bf Symbol}& {\bf Short description} & {\bf Equation} \\
\hline
$L$& Dimensionless parameter relating $r_0$ and $\sqrt{\alpha^\prime}$ & 
(50)\cr
$g_f^2$& 't Hooft coupling constant & (54)\cr
$N$& Number of stacked $p$--branes & (54)\cr
$c_p$& Number relating $g_s$ to the field theory coupling constant& (54)\cr
$k$& Factor denoting the scaling of the $p$--brane tension with $g_s$ & (54)\cr
$x$& Number giving the length dimension of $g_f^2$ & (54)\cr
$\tau_D$& Effective string tension in the dual frame & (61)\cr
$g_{eff}^2$& Effective dimensionless coupling constant & (62)\cr
$U$& D$p$--brane energy scale of fixed stretched string lengths & (63)\cr
\hline
\end{tabular}
\end{center}
\bigskip
}
\newpage
\item {\bf Section 5}
{\small
\begin{center}
\begin{tabular}{|c|c|c|}
\hline
{\bf Symbol}& {\bf Short description} & {\bf Equation} \\
\hline
${\cal L}$& Quantum mechanics Lagrangian & (90)\cr
$m$& Mass of particle probing background geometry & (90)\cr
$q$& Charge of particle probing background geometry & (90)\cr
${\cal H}$& Quantum mechanics Hamiltonian & (92)\cr
$f$& Function in quantum mechanics Hamiltonian & (93)\cr
$g$& Potential in quantum mechanics Hamiltonian & (93)\cr
${\cal V}$& Function denoting the deformation from `classical' QM & (96)\cr
${\cal K}$& Generator of special conformal transformations & (98)\cr
${\cal D}$& Generator of dilatation transformations & (98)\cr
\hline
\end{tabular}
\end{center}
\bigskip
}
\end{itemize}

\newpage

%%%%%%%%%%%%%%%%%%%%%%%%%%%%%%%%%%%%%%%%%%%%%%%%%%%%%%%%%%%

\end{document}